\documentclass[sigconf,10 pt]{acmart}

\usepackage{url}
\AtBeginDocument{%
  \providecommand\BibTeX{{%
    \normalfont B\kern-0.5em{\scshape i\kern-0.25em b}\kern-0.8em\TeX}}}

\copyrightyear{2023} 
\acmYear{2023} 
\setcopyright{acmlicensed}
\acmPrice{15.00}
\acmDOI{10.1145/3620678.3624783}
\acmISBN{979-8-4007-0387-4/23/11}

\acmConference[SoCC '23]{ACM Symposium on Cloud Computing}{October 30--November 1, 2023}{Santa Cruz, CA, USA} 
\acmBooktitle{ACM Symposium on Cloud Computing (SoCC '23), October 30--November 1, 2023, Santa Cruz, CA, USA}

\usepackage{graphicx}
\usepackage{tikz}
\usepackage{caption}
\usepackage{subcaption}
\usepackage{circledsteps}
\usepackage{xcolor,colortbl}
\usepackage{multirow}
\usepackage{tabularx}
\usepackage{hyperref}
\usepackage{cleveref}
\usepackage{tcolorbox}
\usepackage{float}
\usepackage{balance}

\definecolor{steporange}{RGB}{255,191,147}
\pgfkeys{/csteps/inner color=black}
\pgfkeys{/csteps/outer color=black}
\pgfkeys{/csteps/fill color=steporange}

\definecolor{lightred}{HTML}{FFF5F5}
\definecolor{lightblue}{HTML}{F5F6FF}
\definecolor{lightgreen}{HTML}{F5FFF7}
\definecolor{yellow2}{HTML}{FFFFEB}

\newenvironment{tightlist}{
\begin{list}{$\bullet$}{
    \setlength{\topsep}{.1em}
    \setlength{\partopsep}{0in}
    \setlength{\parskip}{0in}
    \setlength{\itemsep}{0in}
    \setlength{\parsep}{0in}
    \setlength{\leftmargin}{1em}
    \setlength{\rightmargin}{0in}
    \setlength{\itemindent}{0in}
}}
{\end{list}}

\newcolumntype{r}{>{\columncolor{lightred}}c}
\newcolumntype{b}{>{\columncolor{lightblue}}c}
\newcolumntype{g}{>{\columncolor{lightgreen}}c}
\newcolumntype{p}{>{\columncolor{yellow2}}c}
\usepackage{array}
\newcolumntype{?}{!{\vrule width 1pt}}

\newcolumntype{L}[1]{>{\raggedright\let\newline\\\arraybackslash\hspace{0pt}}m{#1}}
\newcolumntype{C}[1]{>{\centering\let\newline\\\arraybackslash\hspace{0pt}}m{#1}}
\newcolumntype{R}[1]{>{\raggedleft\let\newline\\\arraybackslash\hspace{0pt}}m{#1}}

\begin{document}




\title[How Does It Function?]{How Does It Function? Characterizing Long-term Trends in Production Serverless Workloads}


\author{Artjom Joosen, Ahmed~Hassan, Martin Asenov, Rajkarn Singh, Luke~Darlow}
\affiliation{%
  \institution{Huawei Edinburgh Research~Centre}
  \city{Edinburgh}
  \country{United Kingdom}
}





\author{Jianfeng Wang}
\orcid{0009-0001-4709-1865}
\affiliation{%
  \institution{Hangzhou Research Centre, Central Software Institute, Huawei}
  \city{Hangzhou City}
  \country{China}
}

\author{Adam Barker}
\orcid{0000-0002-0463-7207}
\email{sirlab@huawei.com}
\affiliation{%
  \institution{Huawei Edinburgh Research Centre and School of Computer Science, University of St Andrews}
 \city{}
  \country{United Kingdom}
}

\newcommand{\revision}[1]{#1} 

\renewcommand{\shortauthors}{Joosen, et al.}


\begin{abstract}

This paper releases and analyzes two new Huawei cloud serverless traces. The traces span a period of over 7 months with over 1.4 trillion function invocations combined. The first trace is derived from Huawei’s internal workloads and contains detailed per-second statistics for 200 functions running across multiple Huawei cloud data centers. The second trace is a representative workload from Huawei’s public FaaS platform. This trace contains per-minute arrival rates for over 5000 functions running in a single Huawei data center. 
We present the internals of a production FaaS platform by characterizing resource consumption, cold-start times, programming languages used, periodicity, per-second versus per-minute burstiness, correlations, and popularity. Our findings show that there is considerable diversity in how serverless functions behave: requests vary by up to 9 orders of magnitude across functions, with some functions executed over 1 billion times per day; scheduling time, execution time and cold-start distributions vary across 2 to 4 orders of magnitude and have very long tails; and function invocation counts demonstrate strong periodicity for many individual functions and on an aggregate level. Our analysis also highlights the need for further research in estimating resource reservations and time-series prediction to account for the huge diversity in how serverless functions behave.


\end{abstract}



\maketitle

\section{Introduction}
Serverless computing and Function-as-a-Service (FaaS) provides cloud programmers with a convenient programming paradigm for event-based workloads~\cite{serveless_in_the_wild_2020}. Today, serverless computing is widely used for applications ranging from media processing~\cite{cosmos} to processing requests from vending machines~\cite{cola}. While serverless adoption has been increasing, there are few available studies into how production serverless systems actually perform, characterizing behavior of both users and systems in operation~\cite{serveless_in_the_wild_2020}. However, such an understanding is important for systems research on how to build, manage, and enhance serverless offerings in cloud data centers. In addition, having open datasets allows for new insights into these systems and how they operate. To the best of our knowledge, there is only one existing serverless dataset available for the research community spanning 14 days of Azure function invocations~\cite{AzureGit}. This dataset contains over 70 thousand functions, but provides only a limited number of features, a short duration of 14 days, and a coarse-grained log of per-minute invocations. 
 
In this paper, we analyze and open-source two traces from Huawei serverless systems~\footnote{\label{footnote:traces}Huawei traces available at \url{https://github.com/sir-lab/data-release}}. The first trace is at per-second granularity and consists of 200 functions from private internal Huawei workloads running across multiple data centers with 141 days of data collected over a period of 234 days. This provides insights into fine-grained performance of these systems including response times, cold-start times, and number of invocations per second for each function, as well as average CPU and memory usage per minute for each function. The second trace is a coarse-grained trace of over 5000 functions hosted in one availability zone from one Huawei data center for 26 consecutive days. This second trace provides a coarse-grained understanding of the evolution of the workload on a single availability zone for over 5000 functions.

\begin{table*}[!htbp]
	\centering
    \begin{tabular}{ |c|c|c|c|c| } 
            \hline
            Dataset & Field & Description & Res & Unit \\
            \hline
            \multirow{10}{6em}{Huawei\\Private} & Function ID & Unique function identifier & - & -\\
            & Timestamp & Time when request is received & sec & sec\\
            & Requests & Number of function invocations & sec & - \\ 
            & Function delay & Average execution time averaged over all pods running that function & sec & ms\\ 
            & Platform delay & Average platform overhead averaged over all pods running that function & sec & ms \\ 
            & CPU usage & Percentage of allocated CPU used per function averaged over all pods & min & \% \\ 
            & Mem usage & Percentage of allocated memory used per function averaged over all pods & min & \% \\ 
            & CPU limit & Allocated CPU for all pods running that function (normalised) & min & cores \\ 
            & Mem limit & Allocated memory for all pods running that function & min & MB\\ 
            & Instances & Number of pods allocated to that function & min & -\\ 
            \hline
            \multirow{6}{6em}{Huawei\\Public} & Function ID & Unique function identifier & -& - \\ 
            & Timestamp & Time when request is received & min & min\\
            & Requests &  Number of function invocations & min & -\\
            & Cold start cost* & Cold start times plus some other overheads & sec & ms\\
             & Package sizes* & Size of functions & - & MB\\
             & Language* & Programming language used  & - & -\\
            \hline
            \end{tabular}
	\caption{Summary of our dataset fields with each field's associated time-measurement granularity. The data will be released as an open-source repository following the format in the Table. Note that fields with an asterisk (*) will not be released.}
 \label{tab:datasets}
\end{table*}

This paper reports the first in-depth analysis of these two Huawei production traces, covering statistical features of our workloads, followed by a longitudinal analysis of periodicity and ranking of functions across our traces. We then formulate the prediction of these function invocations as a time series forecasting problem using a \emph{global univariate model} and demonstrate the challenges of forecasting with models of varying complexities on both traces. Our main contributions can be summarized as follows:
\begin{tightlist}
    \item  \textbf{Characterization of production FaaS workloads:~}We provide a first-of-its-kind analysis of two long-term traces (26 days; 141 days spanning 234 days, with gaps) from production serverless systems, together comprising over 5200 functions. Our analysis covers request arrival distributions, cold start, scheduling, network, function execution times, code package size, resource usage and feature-level correlations.\looseness=-1

    \item \textbf{Longitudinal periodicity and ranking analysis:~}We conduct an in-depth longitudinal analysis of periodicity and ranking of functions across our traces. These analyses are of particular importance regarding function invocation, as they demonstrate a paradigm of data that exhibits trends over the long-term (many days), yet which is also fine-grained.

    \item \textbf{Workload forecasting:~}We demonstrate the efficacy of several machine learning models on forecasting function request data, with the intended application of  autoscaling and scheduling. We provide an argument for why global univariate models are better suited (compared to standard univariate or multivariate models) to this type of data. We also identify a challenge resulting from such data: ingesting and forecasting very fine-grained (per second) and long-term (over several days or weeks). We refer to such data as FGLT data. Facing this challenge may require rethinking how modern methods operate. We hope to see novel modeling approaches emerge based on this data and the associated challenge.

    \item \textbf{Open-source trace release:~}We open-source$^1$ both traces to the research community, encouraging further analysis on serverless systems. Our traces provide a rich set of metrics, and with 141 days of data, facilitate longitudinal studies to better understand serverless systems over longer durations.
\end{tightlist}

\vspace{-3mm}
\begin{figure}[!htbp]
	\centering
    \includegraphics[width=1\linewidth]{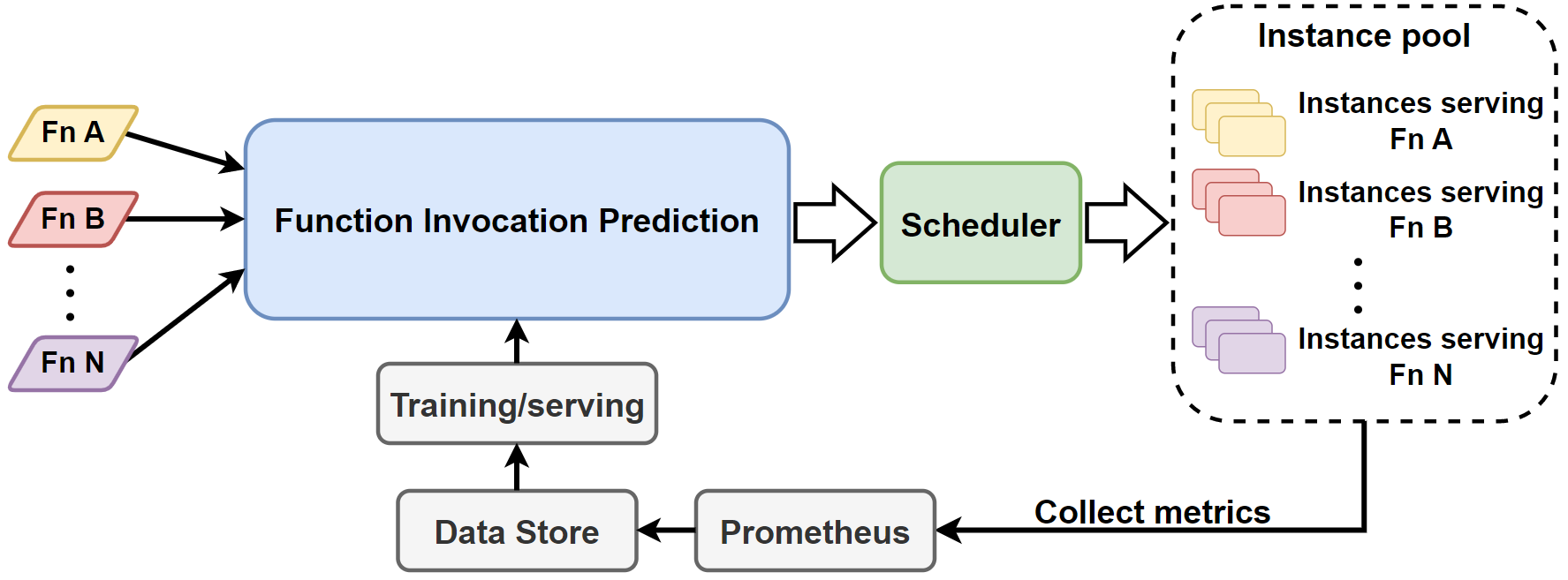}  
    \vspace{- 4 mm}
	\caption{Architecture of our serverless platforms and where our function invocation prediction model will be used.}
    \label{fig:architecture_diagram}
\end{figure}

\section{Overview of our datasets} \label{section:overview}
\label{section:background}

This section describes the lengths and features of our traces as well as granularities and other aggregate characteristics.

\subsection{Serverless at Huawei}

Serverless computing facilitated by Function-as-a-Service (FaaS) platforms allow developers to build, run and deploy event-driven stateless functions without the overhead of having to provision and manage servers or backend infrastructure. This, in turn, allows developers to focus on the functions they develop, rather than how they will run on the cloud~\cite{jangda2019formal,twosteps}. Today, all major cloud providers have serverless offerings. Under the hood, middleware such as Apache OpenWhisk~\cite{openwhisk}, Knative~\cite{knative}, or OpenShift Serverless~\cite{openshift} are used to manage the functions submitted to the system. The functions are then mapped to either containers~\cite{yu2023following} or lightweight virtual machines~\cite{agache2020firecracker}. These containers are then orchestrated into larger applications using cloud orchestration software such as Kubernetes.\looseness=-1

Serverless systems at Huawei are built on top of containers and \emph{YuanRong}, a scalable distributed computing kernel for serverless workloads. YuanRong has been widely deployed across China, Europe and Asia Pacific in nearly 20 availability zones called `regions', and serves tens of thousands of enterprise customers across a diverse range of workloads including data analysis, business IT applications, and deep neural network model training and serving. YuanRong is used to process up to 20 billion requests per day. Figure~\ref{fig:architecture_diagram} shows a simplified architecture of YuanRong. 

\textbf{Function invocation prediction} is the task of predicting the number of function invocations arriving on the platform at a given time for a given function. Predictions produced by a time series forecasting model inform the scheduler and are used to preemptively allocate an appropriate number of pods to serve a particular function in order to reduce the number and time of cold starts. On Huawei platforms, the most popular functions are predicted with powerful models and predictions are made every day for the entire subsequent day. Predictions may be made for every minute or even every second of the subsequent day. These forecasting models use historical data, which is collected by Prometheus, to inform their predictions, as shown in Figure~\ref{fig:architecture_diagram}.

\subsection{Our Datasets}

In this paper, we present and analyze two new datasets from YuanRong's serverless workloads collected using the pipeline in Figure~\ref{fig:architecture_diagram}. These two traces span a period of more than 7 months with over 1.4 trillion function invocations in both datasets combined. The two traces we open-source and analyze in this paper are:

\begin{enumerate}
    \item 
\textbf{Huawei Cloud Private Functions Trace 2023} \newline \textbf{(Huawei Private):} This trace is derived from Huawei's internal workloads. It contains eight different metrics for 200 functions for a period spanning 234 days. Some metrics are reported per second, while others are reported per minute. This trace contains data on the platform itself including CPU and memory usage and limits, along with platform and function delays. 

\item 
\textbf{Huawei Cloud Public Functions Trace 2023} \newline \textbf{(Huawei Public):} This trace is a representative workload from applications running on Huawei's public-facing FaaS platform. Huawei Public contains more functions than Huawei Private but is reported at the minute level and over a shorter duration, totaling 5019 functions over 26 days. We also analyze language types, package sizes, and cold start times for this platform, but do not release these.
\end{enumerate}

\begin{figure}[H]
    \centering
    \begin{subfigure}[b]{0.49\textwidth}
       \includegraphics[width=\linewidth]{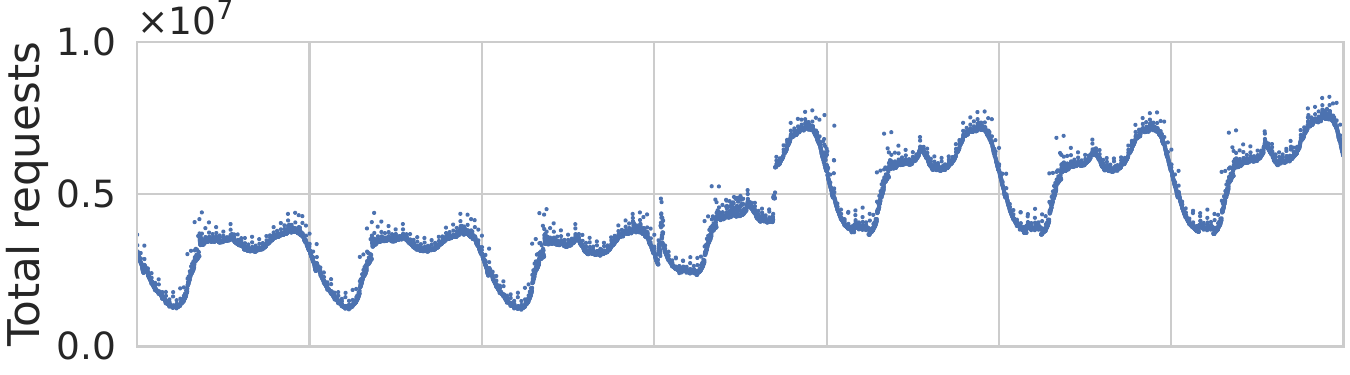}
              \vspace{- 5 mm}
       \caption{Huawei Private.}
       \label{fig:all_requests_time_series_plot_Platform_1} 
    \end{subfigure}
    
    \begin{subfigure}[b]{0.49\textwidth}
       \includegraphics[width=1\linewidth]{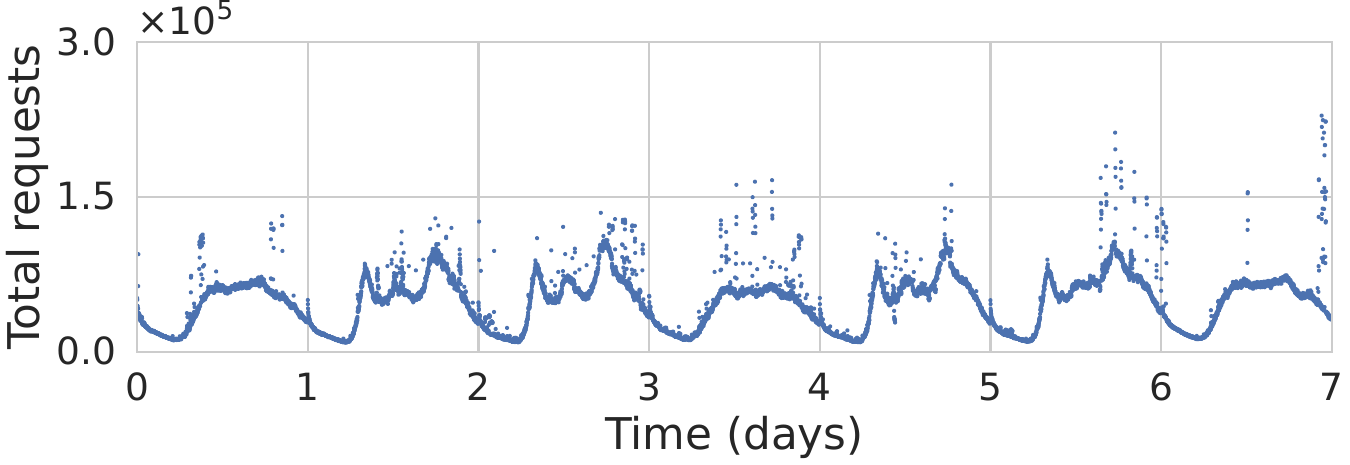}
        \vspace{- 5 mm}
       \caption{Huawei Public.}
    \label{fig:all_requests_time_series_plot_Platform_2}
    \end{subfigure}
    \vspace{- 5 mm}
    \caption{Sum of all requests per minute plotted for the first 7 days of Huawei Private and Huawei Public.}
    \label{fig:all_requests_time_series_plot}
\end{figure}
Table~\ref{tab:datasets} displays a summary of both traces with their respective fields. Figure~\ref{fig:all_requests_time_series_plot} shows the total number of invocations for each of the two traces for a sample week.  
Most of the analysis in this paper is done on the Huawei Private trace since it contains more features. We split our characterization and analysis into three different parts. Section~\ref{section:analysis} focuses on the statistical features of our datasets. Section \ref{section:life} discusses periodicity, rankings, and correlations within and across functions that inform certain design choices for ML approaches in Section \ref{section:forecasting}. Section \ref{section:forecasting} describes how to use the insights from previous sections in order to formulate function invocation prediction as a tractable time series forecasting problem, as well as some challenges that arise when forecasting fine-grain per-second data.



\section{Statistical features of our workloads}\label{section:analysis}

This Section characterizes the various features of our datasets, including invocation counts, memory and CPU usage, number of instances, execution times, cold start times, overhead times, code package size, and runtime languages. We focus on the statistical features of workloads using cumulative distribution functions (CDFs). In CDF plots with \emph{all}, \emph{median}, \emph{min}, and \emph{max} curves; \emph{all} represents the CDF of all values across all functions; while \emph{median}, \emph{min}, and \emph{max} are calculated per function (i.e., a percentile calculated for all functions over the entire duration of the dataset), and the CDF is calculated as a cumulative fraction of the number of functions.

\subsection{Request arrivals}
\begin{figure}[H]
	\centering
	\includegraphics[width=1\linewidth]{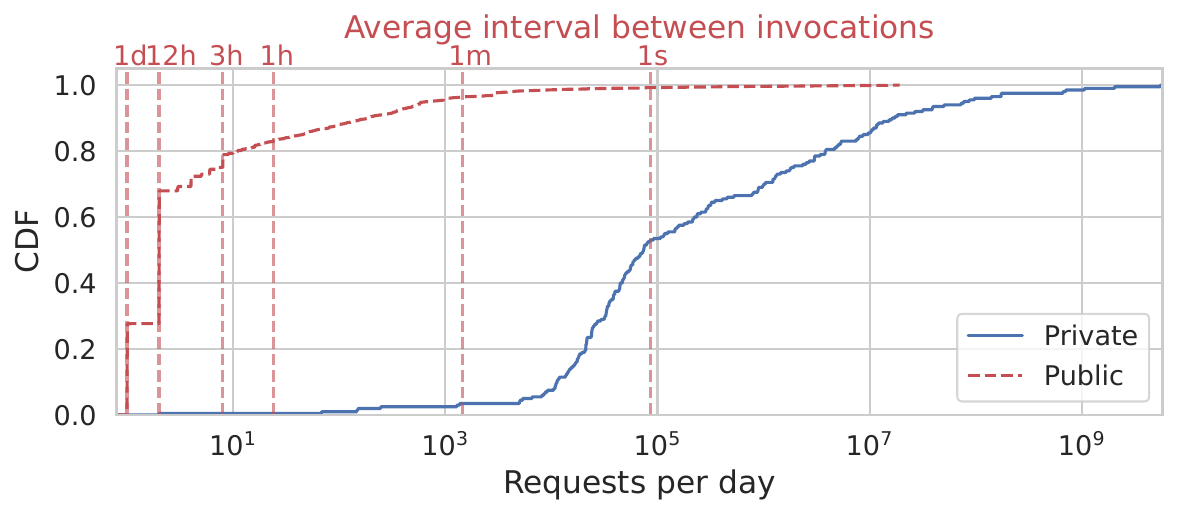}\vspace{-2mm}
	\caption{Requests on a median day per function, with the corresponding average interval between requests for Huawei Private and Huawei Public.}
 \label{fig:cdf_of_functions}
\end{figure}
Figure \ref{fig:cdf_of_functions} shows a CDF of functions by their average number of daily invocations from our two traces. The top horizontal axis shows the mean arrival rate between invocations corresponding to the requests per day on the bottom horizontal axis. To measure requests per day, we compute the median value of each minute of the day across all days in the dataset and take the sum of requests on this median day.
Note that the invocation counts from Huawei Private represent 200 functions of internal Huawei applications using serverless across \emph{all} regions, while Huawei Public contains 5093 functions from \emph{one} region.

Huawei Public has a wide distribution of popularity of functions as measured by requests per day and demonstrates the variety of use within the public FaaS platform. Huawei Private on the other hand has fewer but far more highly requested functions, with some being invoked over 1 billion times per day. Approximately 50\% of functions in Huawei Private are invoked at least once per second, while less than 10\% of functions in Huawei Public are invoked at least once per minute. Thirty percent of functions in Huawei Public are invoked only once per day, and around 70\% of Huawei Public functions are invoked at most once every 12 hours. 
In both datasets, a minority of top functions account for the majority of traffic. This phenomenon is especially present in Huawei Private.\looseness=-1

\begin{figure}
	\centering
	\includegraphics[width=0.85\linewidth]{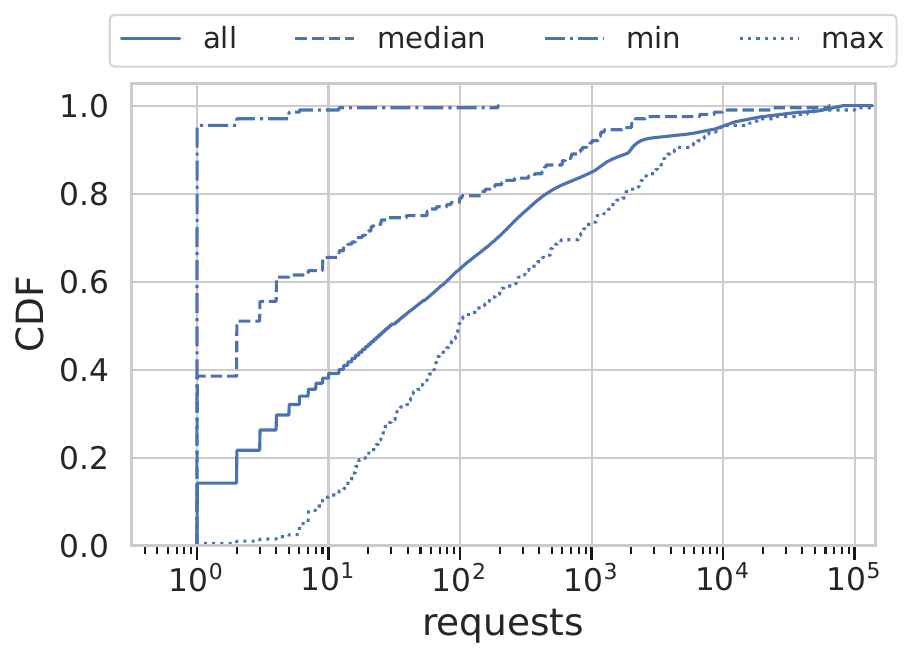}\vspace{-4 mm}
	\caption{Huawei Private request arrivals per-second.}
 \label{fig:cdf_requests}
\end{figure}

To better understand the request arrival rate on a per-second scale, Figure~\ref{fig:cdf_requests} plots the CDF of \emph{all} function invocation counts per second for the Huawei Private dataset. This CDF shows that at least 40\% of all invocation counts have over 100 invocations per-second at any time, with more than 10\% having more than 1000 requests per second. 
In addition to the arrival rate distribution, we plot the distributions of the median, min, and max arrival rates for each of the Huawei Private functions. These distributions can be helpful when designing schedulers using our dataset as they can enable new scheduling approaches that look at the probability of worst and best-case arrivals of requests at any given time.

\begin{tcolorbox}
The number of requests per day varies by nine orders of magnitude across functions in Huawei Private and seven orders of magnitude in Huawei Public. Huawei Public has a similar profile to Azure~\cite {serveless_in_the_wild_2020}, most likely because both are public serverless traces, while Huawei Private has much larger invocation counts.
\end{tcolorbox}

\begin{figure*}
	\centering
 \begin{subfigure}[b]{0.23\linewidth}
	\centering
	\includegraphics[height=3cm]{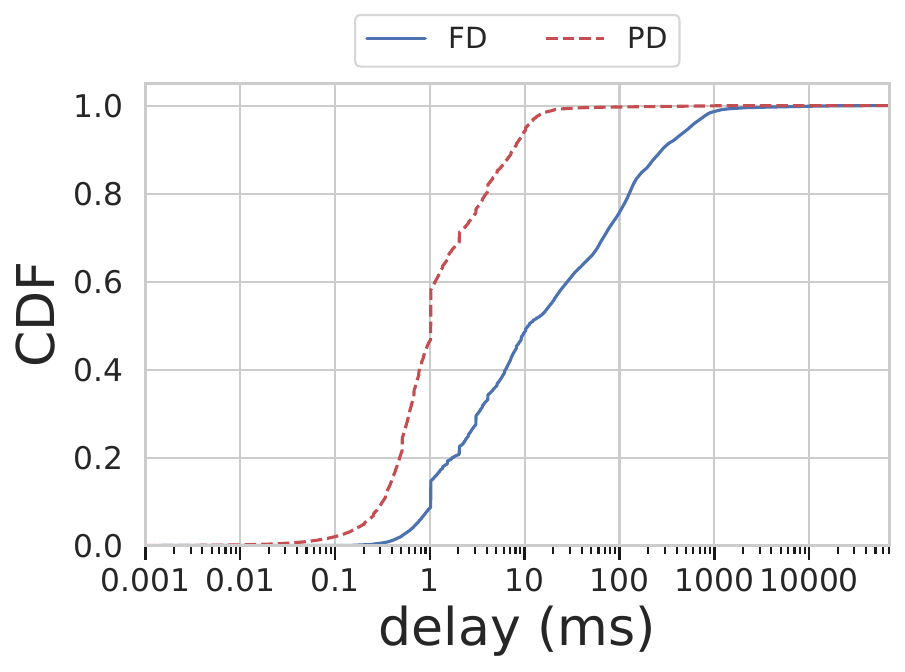}\vspace{- 1 mm}
	\caption{Function Delay, Platform Delay in Huawei Private.}
 \label{fig:cdf_platform_delay}
\end{subfigure}
\hfill
 \begin{subfigure}[b]{0.23\linewidth}
    \includegraphics[height=3cm]{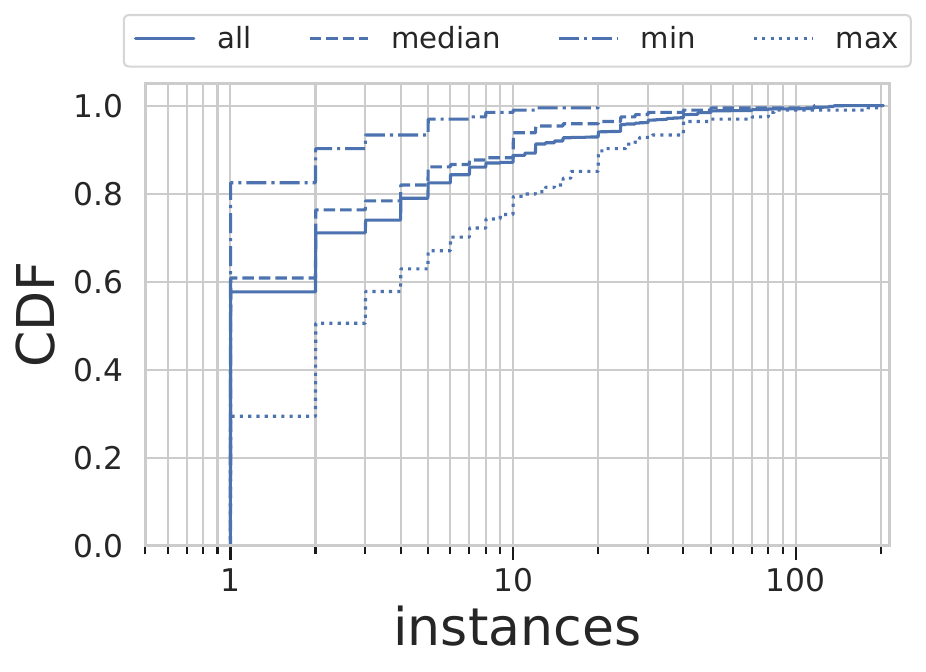}\vspace{-1 mm}
	\caption{Number of allocated instances in Huawei Private.}
 \label{fig:cdf_instances} 
 \end{subfigure}
 \hfill
 \begin{subfigure}[b]{0.23\linewidth}
	\centering
	\includegraphics[height=3cm]{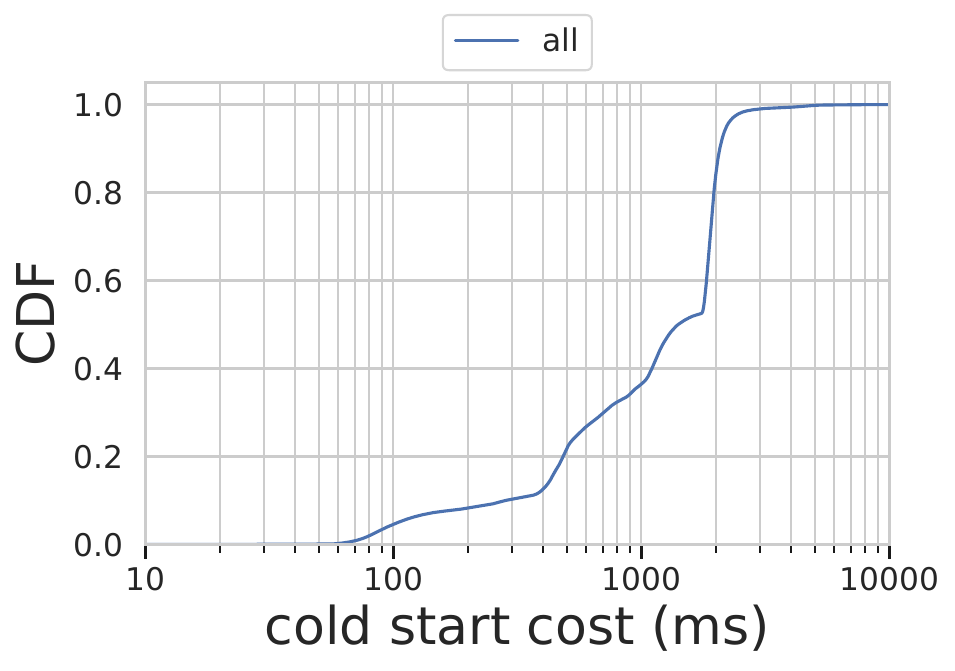}\vspace{- 1mm}
	\caption{Cold start costs measured at Huawei Public.}
 \label{fig:cdf_cold_start}
\end{subfigure}
\hfill
 \begin{subfigure}[b]{0.23\linewidth}
	\centering
	\includegraphics[height=3cm]{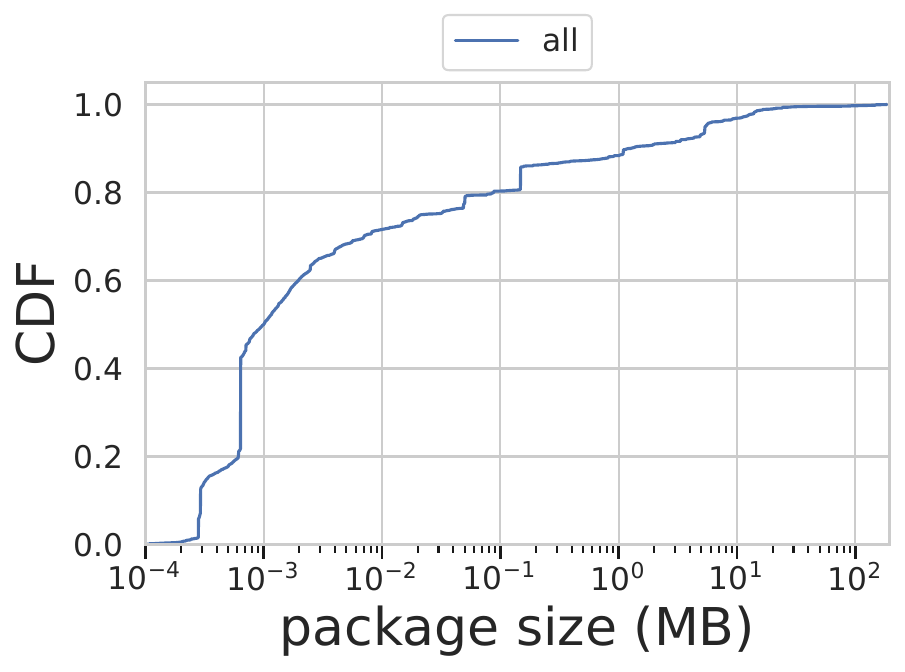}\vspace{-1 mm}
	\caption{Package size per function in Huawei Public.}
 \label{fig:cdf_package_size}
\end{subfigure}
\vspace{- 1 mm}
\caption{The user-perceived times from serverless can be dominated by the platform delay (which includes the scheduling and network setup), by the execution times, or by the cold-starts suggesting that these three metrics need to be optimized. The number of instances and the package sizes can also affect cold starts. }
\end{figure*}

\subsection{Execution time, platform delay, cold-start, and instances}

One potential bottleneck with serverless systems is \emph{cold-starts}~\cite{thomas2020particle,daw2020xanadu}. 
Due to the burstiness of serverless workloads, the system may not have enough containers already provisioned to service incoming request bursts~\cite{li2022help,wang2018peeking}. The system, therefore, needs to start new containers to accommodate these bursts, which adds latency and degrades application performance. This cold-start time is affected by the network cold-start time~\cite{thomas2020particle}, the sizes of the containers, and the need to pull containers from registries~\cite{wang2021faasnet}, among other costs that can affect the total cold-start.

\textbf{Huawei Private delays analysis.} For Huawei Private, we analyze two delays, namely, \textbf{function delay} and \textbf{platform delay}. Function delay is the function execution time measured by our system, i.e., the time it takes a function to complete its task once scheduled. Platform delay includes both scheduling and network delays.  Platform delay has previously been found to dominate user-perceived cold-start times in ``burst-parallel'' serverless jobs~\cite{thomas2020particle}. Burst-parallel serverless jobs are generated by applications that invoke thousands of short-lived distributed functions to complete complex jobs requiring the system to start a large number of containers that requires interconnections~\cite{thomas2020particle, wukang}. Since these burst-parallel functions dominate our Huawei Private data sets, and since the Huawei Private functions represent real applications used by millions of customers using internal services, we analyze these two delays for the Huawei Private dataset.

\begin{figure}[h]
	\centering
	\includegraphics[width=0.8\linewidth]{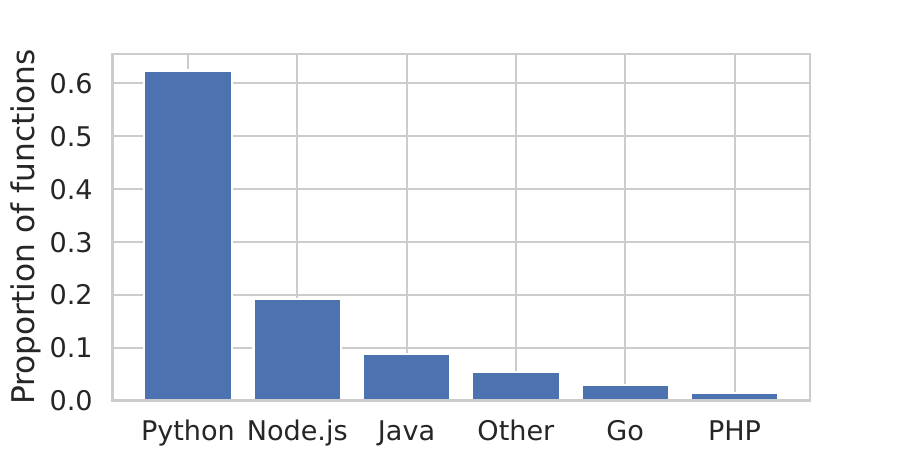}\vspace{-1 mm}
	\caption{Proportion of functions by runtime language in Huawei Public.}
 \label{fig:bar_chart_languages}
\end{figure}

Figure~\ref{fig:cdf_platform_delay} shows the CDFs of both platform and function delays for Huawei Private. Over 60\% of function requests have a platform delay of less than 1ms. Almost all platform delays are less than 10ms. Confirming previous research~\cite{thomas2020particle}, the platform delays have a very long tail due to the burst-parallel nature of the workloads. Next to platform delay, we also plot function delay on the same figure. We see that 50\% of functions execute in less than 10ms, and almost all functions complete within 1 second.

To better understand the long tails in platform delay, we plot the number of instances allocated per function per-minute (solid blue labeled \emph{all}) for Huawei Private in Figure~\ref{fig:cdf_instances} as the number of instances can strongly influence the network-startup times~\cite{thomas2020particle}. Even though a significant percentage of our workload has at least 100 invocations per second (roughly 60\%), we allocate only one instance to serve 60\% of the functions, with only less than 1\% of our allocations being above 100 instances. We believe that these larger allocations are one reason for the longer platform delays as for many functions, the platform delay and the number of instances are correlated as we show later.

\textbf{Huawei Public delays analysis.} Huawei Public represents external workloads deployed by our customers, typically with much fewer insights from our teams on what is actually running. Hence, for Huawei Public, we decided to analyze the cold-start costs which measure the total time for a function to start. Figure~\ref{fig:cdf_cold_start} shows the cold start cost for Huawei Public. We see that approximately 40\% of cold starts take less than 1 second, while over 90\% of requests of cold starts take less than 2 seconds. Major influences on cold-start times include package sizes, memory usage, and language runtimes used~\cite{wang2018peeking}. Figure~\ref{fig:cdf_package_size} shows the package sizes for the functions in Huawei Public which is an indication of the size of the code running, while Figure~\ref{fig:bar_chart_languages} shows the main languages used for implementing these functions. Our data analysis shows that package sizes, the language of implementation, and memory allocation are the three most decisive factors on the cold-start time~\cite{wang2018peeking}. 

\vspace{-2 mm}

\begin{tcolorbox}
Platform delays, execution time, and cold-start time distributions vary considerably and have very long tails. This indicates the importance of optimizing all three delays when building serverless platforms. However, this is especially challenging given the diversity of runtime languages used and their associated performance overheads.
\end{tcolorbox}

\subsection{Resource Usage}

\begin{figure*}
	\centering
 \begin{subfigure}[b]{0.23\textwidth}
    \centering
	\includegraphics[height=2.96cm]{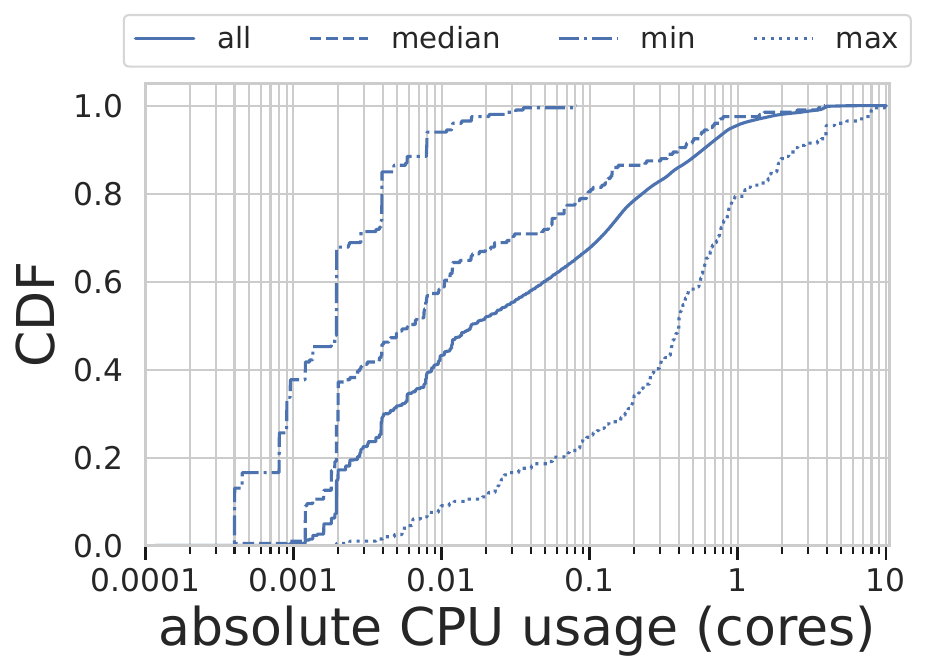}\vspace{-1mm}
	\caption{CPU usage.}
 \label{fig:cdf_absolute_cpu_usage}
 \end{subfigure}
 \hfill
 \begin{subfigure}[b]{0.23\textwidth}
	\centering
	\includegraphics[height=2.96cm]{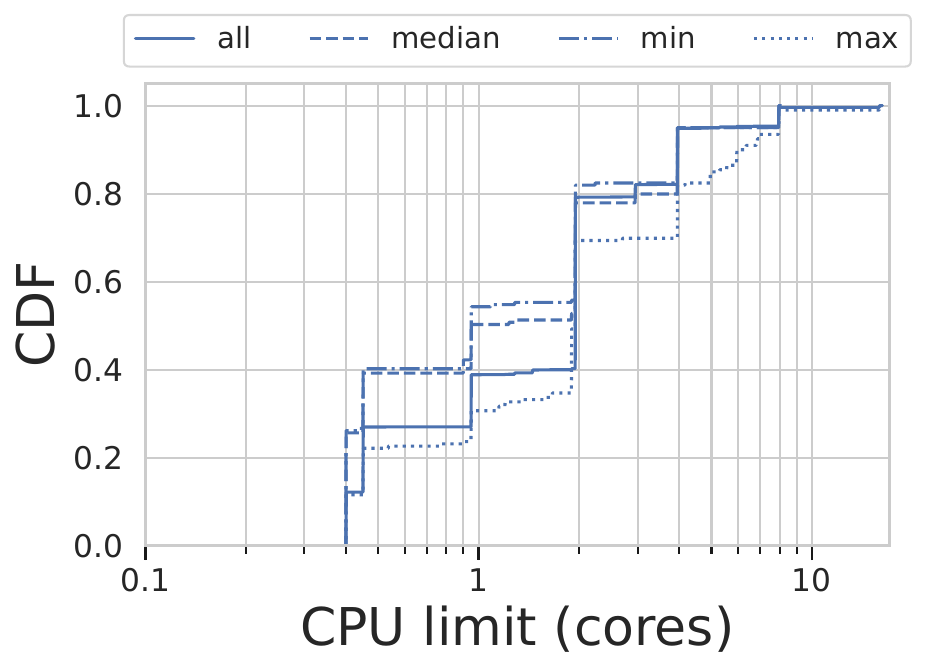}\vspace{-1mm}
	\caption{CPU limit.}
 \label{fig:cdf_cpu_limit}
 \end{subfigure}
\hfill
 \begin{subfigure}[b]{0.23\textwidth}
	\centering
	\includegraphics[height=2.96cm]{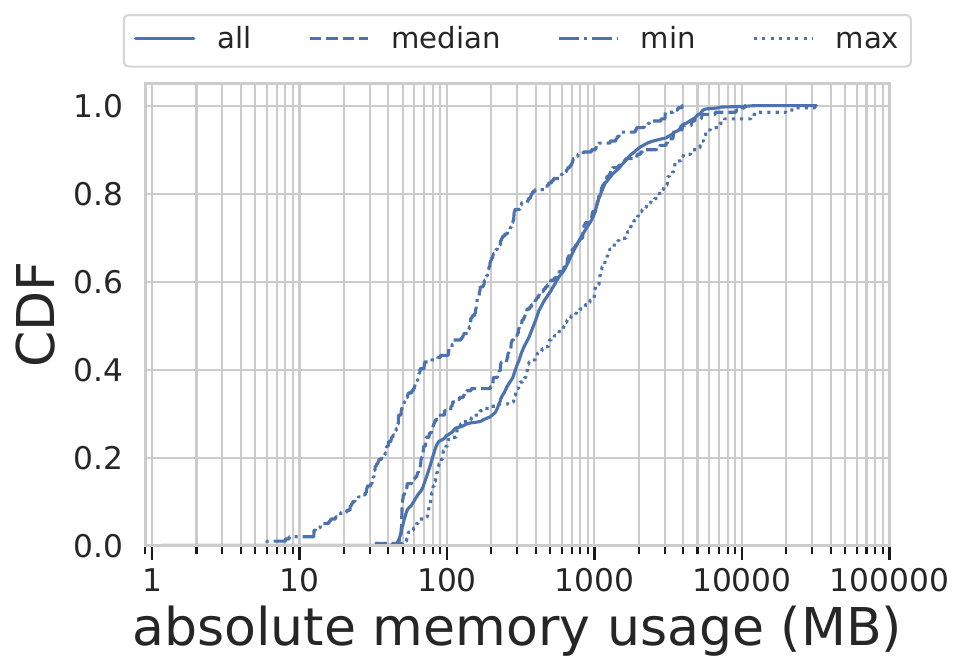}\vspace{-1mm}
	\caption{Memory usage.}
 \label{fig:cdf_absolute_memory_usage}
\end{subfigure}
\hfill
 \begin{subfigure}[b]{0.23\textwidth}
	\centering
	\includegraphics[height=2.96cm]{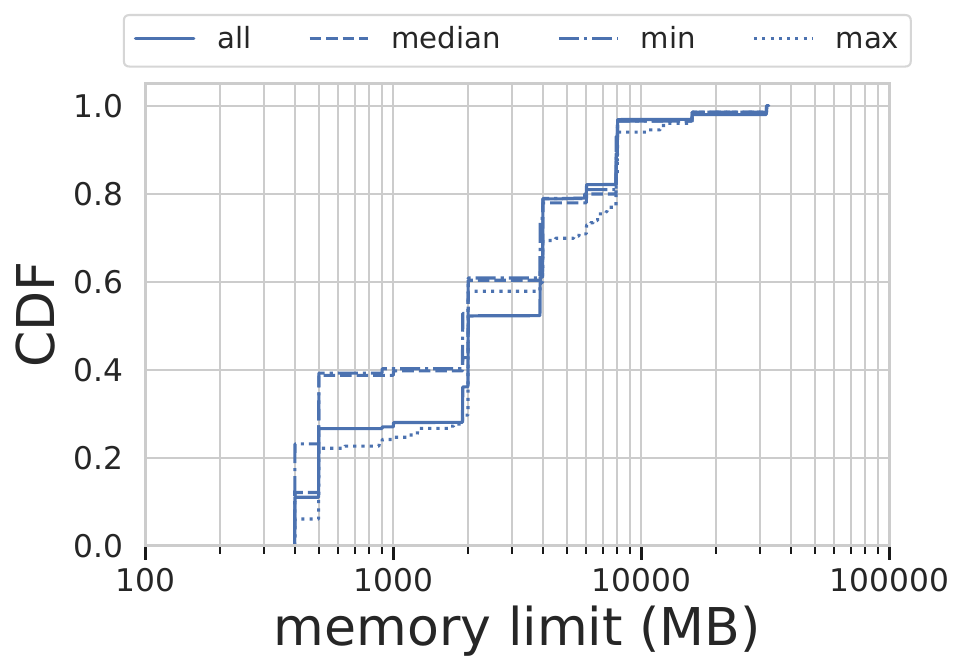}\vspace{-1mm}
	\caption{Memory limit.}
 \label{fig:cdf_memory_limit}
 \end{subfigure}
 \vspace{-4mm}
 \caption{Distribution of resource usage and limit in Huawei Private.}
 \label{fig:cpu_mem_usage_limits}
\end{figure*}

In serverless systems, each function is typically assigned a resource limit by the user that represents the maximum amount of CPU and memory that the function can use. However, functions do not necessarily use all of their assigned resources. To better understand usage versus allocations, we focus on data from Huawei Private. Figure~\ref{fig:cpu_mem_usage_limits} shows the CDFs of absolute CPU usage, CPU limit, absolute memory usage, and memory limit, for all allocations (marked \emph{all}) as well as median, maximum, and minimum for each function. 
Our first observation is that most user-specified limits are much higher than the actual usage, with around 60\% of all allocations using less than 0.1 cores but asking for a limit of more than 1 core. Looking at memory usage in Figure~\ref{fig:cdf_absolute_memory_usage} versus memory limits in Figure~\ref{fig:cdf_memory_limit}, we see a similar trend, with around 50\% of the functions having a limit of around 2 GB, but using only around 400 MB. 
This shows how users of serverless functions are very conservative with their resource requirements. This gives the operators the possibility to reuse some of this slack by over-committing the resources using intelligent scheduling~\cite{verma2015large}. 

When over-committing resources, the scheduler needs to consider the worst-case scenarios of usage, which can be calculated from the max curves for CPU and memory usage in Figures \ref{fig:cdf_absolute_cpu_usage} and \ref{fig:cdf_absolute_memory_usage}. Over-provisioning by users is less conservative when considering the worst-case scenario, e.g. 10\% of functions at some time used 70\% or more of their allocated memory. Hence, scheduler over-commitment must reduce the risk of aggressive over-commitments, especially considering that memory over-allocation can result in failures. There is still room for over-committing the resources while not reducing the quality of service.

\begin{tcolorbox}
Estimating the size of resources required by a function is difficult for many serverless users, leading to over-provisioned resource reservations. This suggests the need for improved and automated approaches for users to determine resource limits, as well as resource reclamation mechanisms that enable serverless operators to reclaim some of these unused resources.
\end{tcolorbox}

\section{The Life of a Function}\label{section:life}
We now focus on individual function behaviors and cross-function relationships. 
We investigate the following characteristics: periodicity, popularity ranking over time, correlations between top-ranking functions, and differences in burstiness between per-second data and aggregated per-minute data. We also compute and discuss correlations between different features. 


\begin{figure}
     \centering
     \begin{subfigure}[b]{0.22\textwidth}
         \centering
         \includegraphics[height=3.5cm]{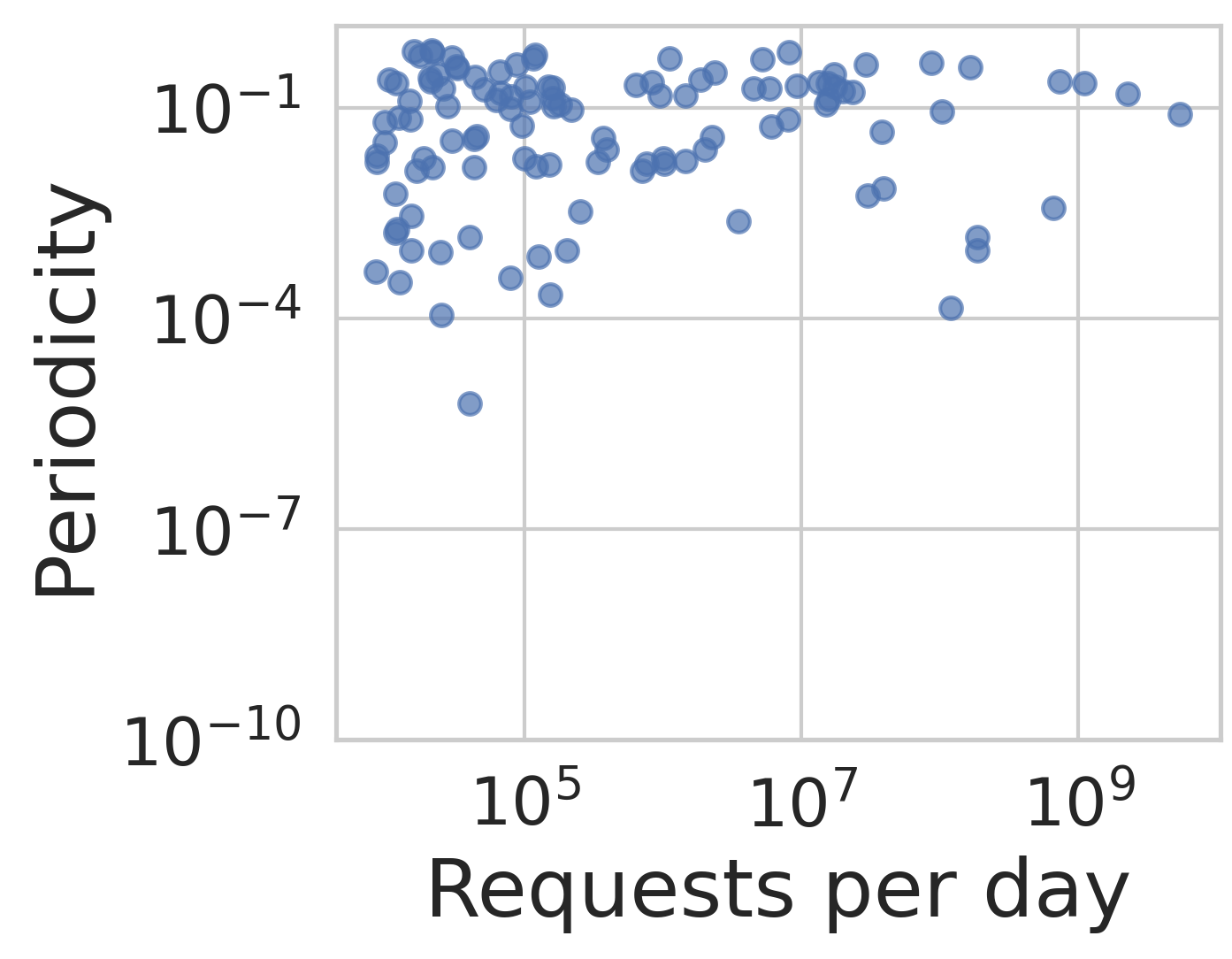}     
         \caption{Huawei Private.}
        \label{fig:periodicity_vs_requests_periodicity_power_no_cutoff_vs_sum_median_avg_day_Platform1}
     \end{subfigure}
     \hfill
     \begin{subfigure}[b]{0.22\textwidth}
         \centering
         \includegraphics[height=3.5cm]{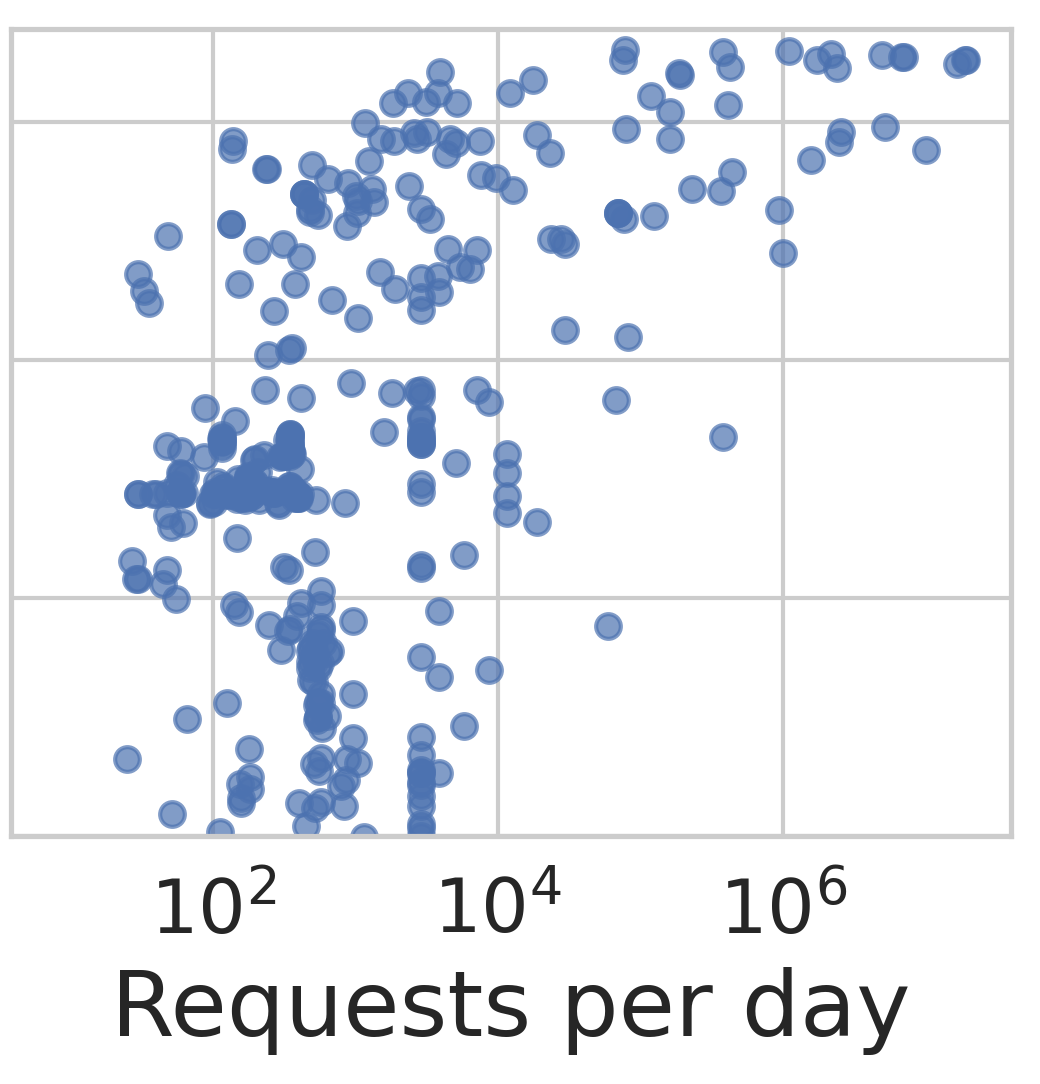}    
         \caption{Huawei Public.}
        \label{fig:periodicity_vs_requests_periodicity_power_no_cutoff_vs_sum_median_avg_day_Platform2}
     \end{subfigure}
     \vspace{-3mm}
    \caption{Periodicity vs number of requests on median day. Periodicity of Huawei Private is generally higher than that Huawei Public. Note that both traces have similar trends for periodicity greater than $10^{-4}$, but Huawei Public has more low-request functions with insignificant periodicities.}
        \label{fig:periodicity_vs_requests_periodicity_power_vs_sum_median_avg_day}
\end{figure}

\subsection{Are functions invocation counts periodic?}

One key characteristic of server workload data is periodicity. This periodicity is seen in many server workloads~\cite{yu2006understanding,atikoglu2012workload,shen2016follow}, including serverless workloads~\cite{serveless_in_the_wild_2020,li2022help}. Identifying the meaningful periodicities will inform scheduling and autoscaling design choices such as the architecture of the model used for forecasting these traces, how often to invoke it to make a prediction, and how long into the future it should predict.\looseness=-1

\textbf{Daily Periodicity.} We analyze the workloads' daily periodicity of each function in the traces. We follow this with a comprehensive analysis of periodicity at any period. The daily periodicity of a function can be measured by the normalized peak height of its power spectral density (PSD) at the daily frequency. PSD is a measure of a time-series' power content distribution with frequency. We measure the PSD value for daily periodicity of all functions and plot them in Figure \ref{fig:periodicity_vs_requests_periodicity_power_vs_sum_median_avg_day} as a scatter plot. Most highly requested functions on each platform have daily periodicity, especially in Huawei Public. Functions with fewer requests per day may also have strong periodicity. There are a few high-request, low-periodicity functions in both datasets, pointing to functions with large, aperiodic bursts. Diurnal patterns are typically due to either human activity during the day, or increased batch processing and analytics jobs during the night.

\begin{figure}
     \centering
     \begin{subfigure}[b]{0.22\textwidth}
         \centering
         \includegraphics[height=6cm]{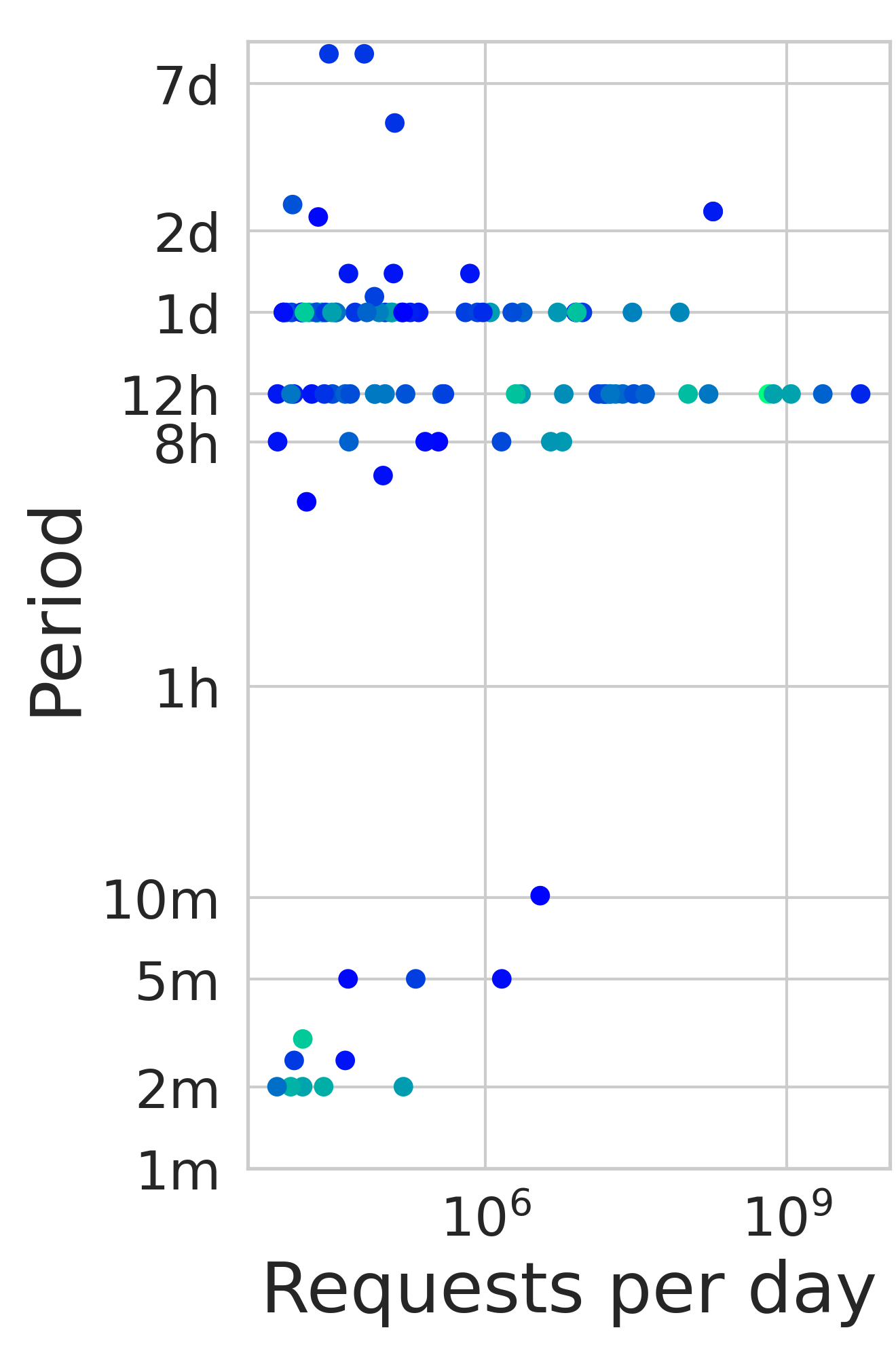}
       \vspace{- 4 mm}
         \caption{Huawei Private.}
         \label{fig:periodicity_vs_requests_max_periodicity_period_vs_sum_median_avg_day_Platform1}
     \end{subfigure}
     \hfill
     \begin{subfigure}[b]{0.22\textwidth}
         \centering
         \includegraphics[height=6cm]{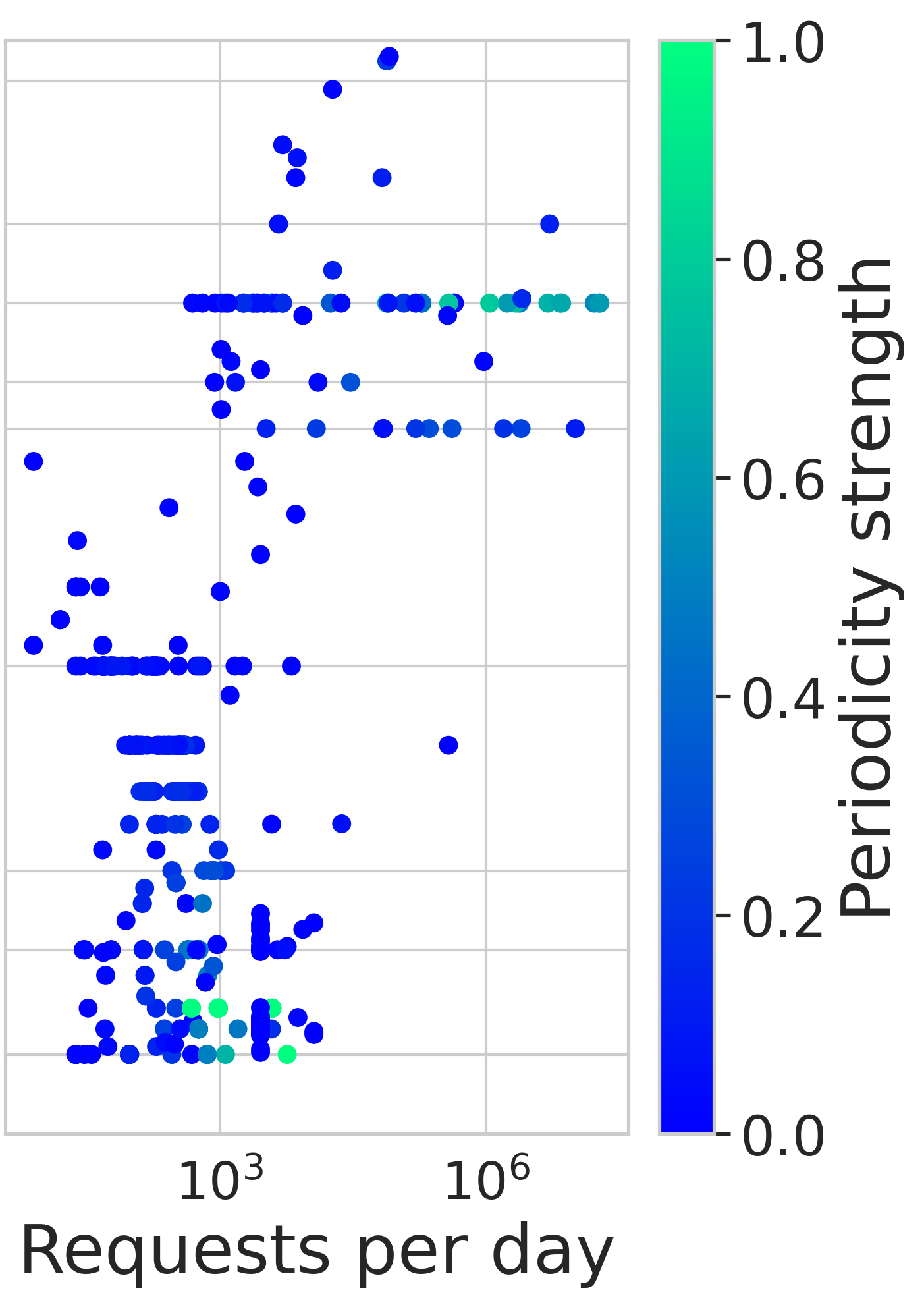}
         \vspace{- 4 mm}
         \caption{Huawei Public.}
         \label{fig:periodicity_vs_requests_max_periodicity_period_vs_sum_median_avg_day_Platform2}
     \end{subfigure}
              \vspace{-1mm}
        \caption{Plot of periodicity of different functions. The horizontal axis is the median number of requests received by a function per day. On the vertical axis is the most prominent periodicity of that function (frequency at which PSD is highest). The color is the height of the largest PSD peak in the normalized PSD (at the aforementioned frequency). Many functions have their most significant periods at 1 day and 12-hour periodicities, including some of the most popular functions.}
        \label{fig:periodicity_vs_requests_max_periodicity_period_vs_sum_median_avg_day}
\end{figure}

\textbf{Other periodicities.} To obtain a more complete picture of periodicity in our datasets, we now examine all periodicities. Given a signal's frequency spectrum, the most significant periodicity is defined by the height of the largest peak in the normalized PSD. The frequency at which this peak occurs represents the period with the strongest influence on the overall signal. We can plot requests per day against the frequency where the PSD is largest, as shown in Figure \ref{fig:periodicity_vs_requests_max_periodicity_period_vs_sum_median_avg_day}. We can then color the points based on the height of the peak, with 1 representing a signal perfectly periodic at that frequency, and 0 representing a completely aperiodic signal. 
In this analysis, we only include functions with more than one consecutive day of data since a large number of functions in our workloads are invoked for less than one day. 

Figure \ref{fig:periodicity_vs_requests_max_periodicity_period_vs_sum_median_avg_day} shows the relationship between invocations per day for each function, the interval at which function invocations have maximum periodicity, and the power of that periodicity from PSD analysis, i.e., the frequency with highest power for each function given its invocation rates. We see significant periodicities at 1 day, 12 hours, 8 hours, and smaller minute-level intervals. This is an interesting observation for serverless system operators. While 24, 12, and 8-hour intervals for strong periodicity can be explained by the day-and-night effect, the per-minute periodicity cannot. 

\begin{figure}
    \centering
    \begin{subfigure}[b]{0.23\textwidth}
       \includegraphics[height=2.04cm]{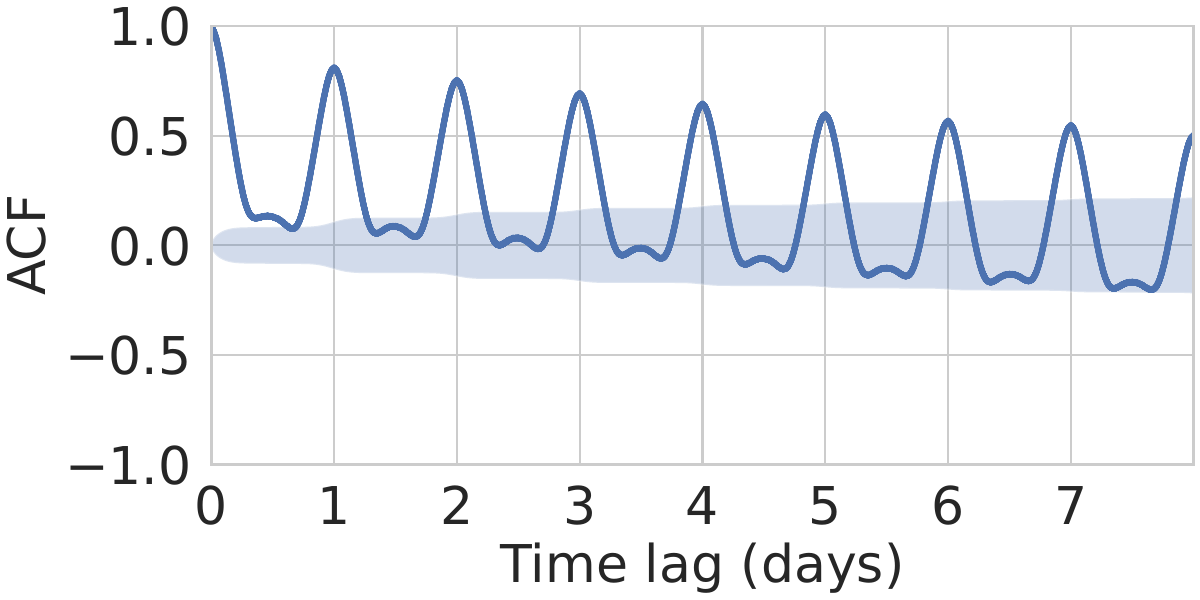}
       \caption{Huawei Private.}
       \label{fig:all_requests_acf_plot_Platform_1} 
    \end{subfigure}
    \begin{subfigure}[b]{0.23\textwidth}
       \includegraphics[height=1.96cm]{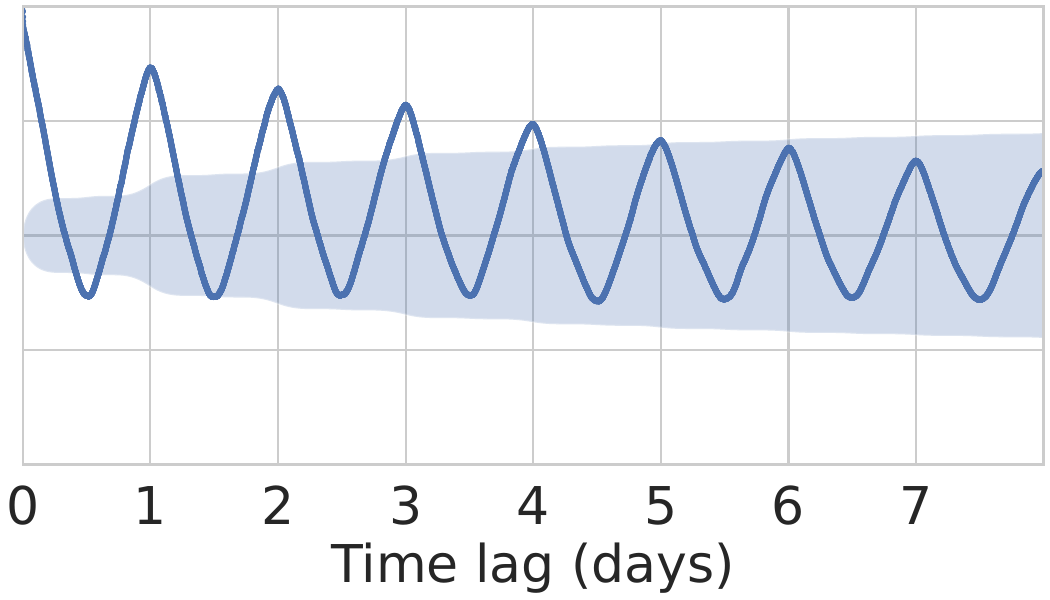}
       \caption{Huawei Public.}
       \label{fig:all_requests_acf_plot_Platform_2}
    \end{subfigure}
         \vspace{- 3 mm}
    \caption{Correlogram of sum of all requests for entire duration against time lag in minutes. The Bartlett confidence intervals are also shown. If autocorrelation is within the confidence interval, it is not significant.}
    \label{fig:correlogram}
    \vspace{-3mm}
\end{figure}

\textbf{Aggregate periodicity.} To understand the overall periodic behavior of workloads, we study the autocorrelation function (ACF) of the total requests for both platforms. At a given lag $\tau$, autocorrelation computes the correlation of a signal with that same signal, but with the latter copy shifted by $\tau$. The ACF is then plotted against lag to visualize the signal's self-similarity when shifted by that lag. Figure~\ref{fig:correlogram} shows the ACF with lags up to eight days. Both datasets have peaks in their autocorrelation at integer multiples of one day. Other than small lags of one minute or one hour, autocorrelation is highest at a lag of exactly one day, underscoring the significance of daily periodicity across functions. Comparing the two datasets, Huawei Private has a marginally stronger autocorrelation at a lag of one day. However, Huawei Public has a consistently larger confidence interval. If the correlation is within the confidence interval, then it is not statistically significant. Also note that Huawei Private has minor peaks at half-day lags, where Huawei Public has troughs, which aligns with the greater number of points on the 12h line in Figure \ref{fig:periodicity_vs_requests_max_periodicity_period_vs_sum_median_avg_day_Platform1} than in Figure \ref{fig:periodicity_vs_requests_max_periodicity_period_vs_sum_median_avg_day_Platform2}.

\begin{figure}
	\centering
	\includegraphics[width=1\linewidth]{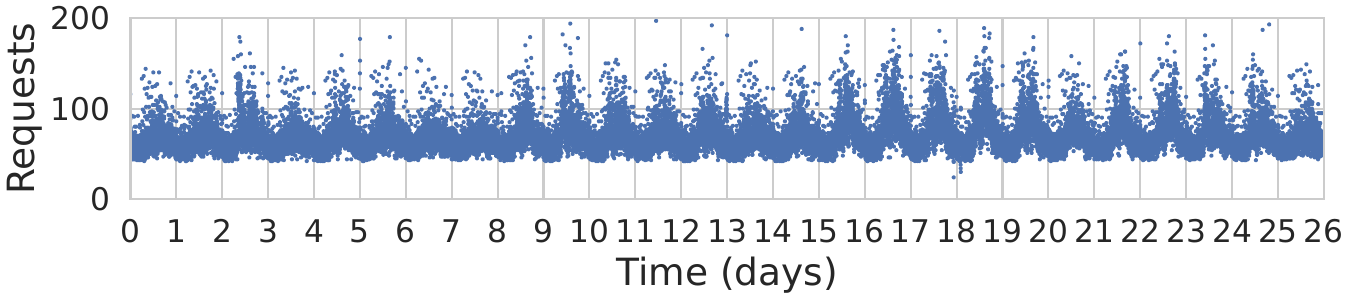}
	\caption{Sum of requests for the 3000 least popular functions in Huawei Public.}
 \label{fig:bottom_3000_requests_plot_Platform_2}
\end{figure}

In addition to the top functions and the total number of requests being periodic, we have found that the least popular functions together also exhibit periodicity. This phenomenon is especially significant for Huawei Public, where the bottom 3000 functions together show clear daily periodicity over all 26 days, as shown in Figure~\ref{fig:bottom_3000_requests_plot_Platform_2}. This is interesting from the perspective of scheduling and bin packing groups of less popular functions together, which are likely to exhibit higher cold start times. 

\begin{tcolorbox}
    A significant number of functions in both platforms have strong periodicity, especially at daily frequencies or frequencies that divide equally into one day, such as 8 or 12 hours. Periodicity is especially significant for more highly requested functions, as well as the aggregate number of requests on our platforms.
\end{tcolorbox}

\subsection{Function popularity over time} \label{section:ranking_analysis}

An important aspect to consider is function popularity as the most popular functions constitute a large portion of the traffic on our platforms. Hence, it is important to understand how the ranking of the top functions changes over time.  

\begin{figure}
    \centering
    \begin{subfigure}[b]{0.49\textwidth}
       \includegraphics[width=0.9\linewidth]{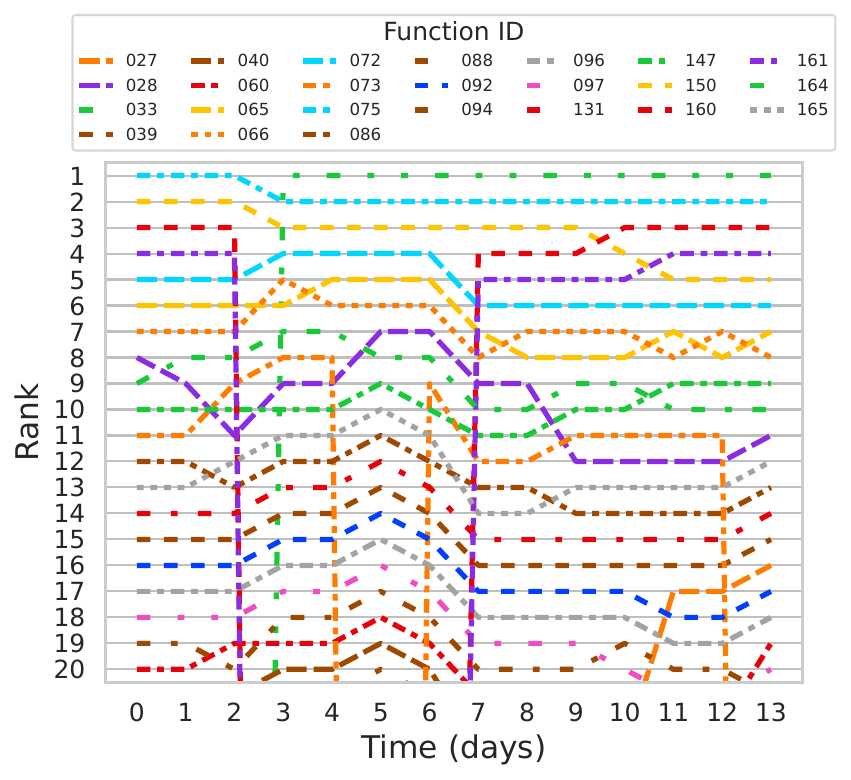}
                     \vspace{- 2 mm}
       \caption{Huawei Private.}
       \label{fig:bump_chart_wf}
    \end{subfigure}
    \begin{subfigure}[b]{0.49\textwidth}
       \includegraphics[width=0.9\linewidth]{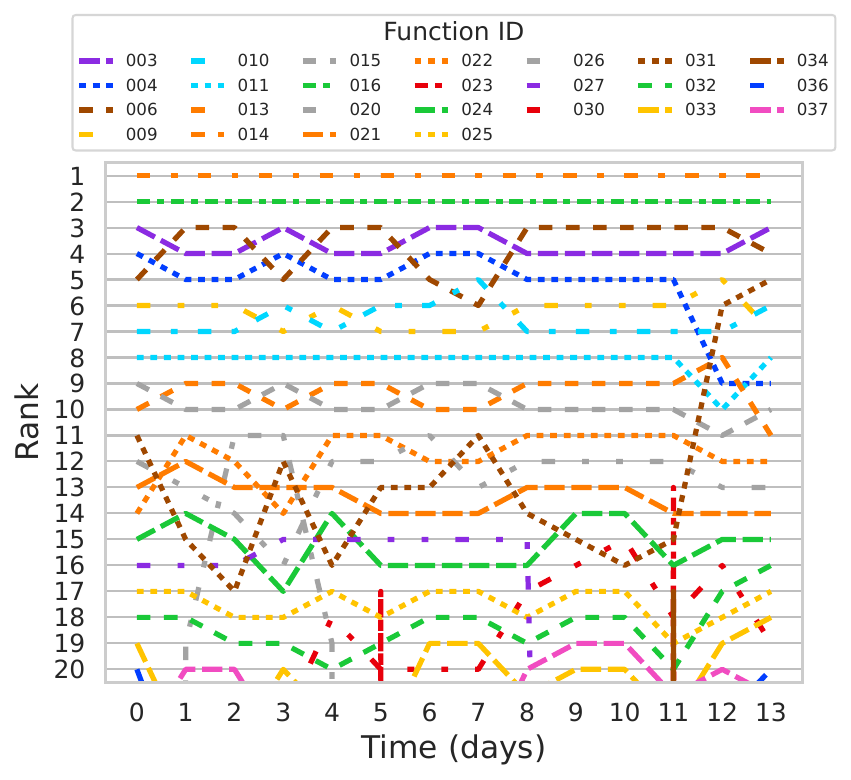}
                     \vspace{- 2 mm}
       \caption{Huawei Public.}
       \label{fig:bump_chart_fg} 
    \end{subfigure}
     \vspace{- 4 mm}
    \caption{Bump chart visualizing the functions that occupy the top 20 places in the daily ranking of functions by their median number of requests on that day.}
    \label{fig:bump_charts}
\end{figure}

The ranking of the top functions rarely remains the same for long. To best visualize this dynamic behavior, we show \emph{bump charts} of function ranking for two weeks of each trace. A bump chart shows the ranking on the vertical axis, and how the ranks of these functions change from day to day. Figure~\ref{fig:bump_chart_wf} shows the bump chart for Huawei Private and Figure~\ref{fig:bump_chart_fg} shows the bump chart for Huawei Public. Our first observation is that 25 unique functions appear in the top 20 rankings over two weeks for Huawei Public, while 24 appear in the same rankings for Huawei Private, suggesting that there is little change in rankings in the top ranks. Some of these functions appear only briefly.

\begin{figure}
    \centering
    \begin{subfigure}[b]{0.49\textwidth}
       \includegraphics[width=1\linewidth]{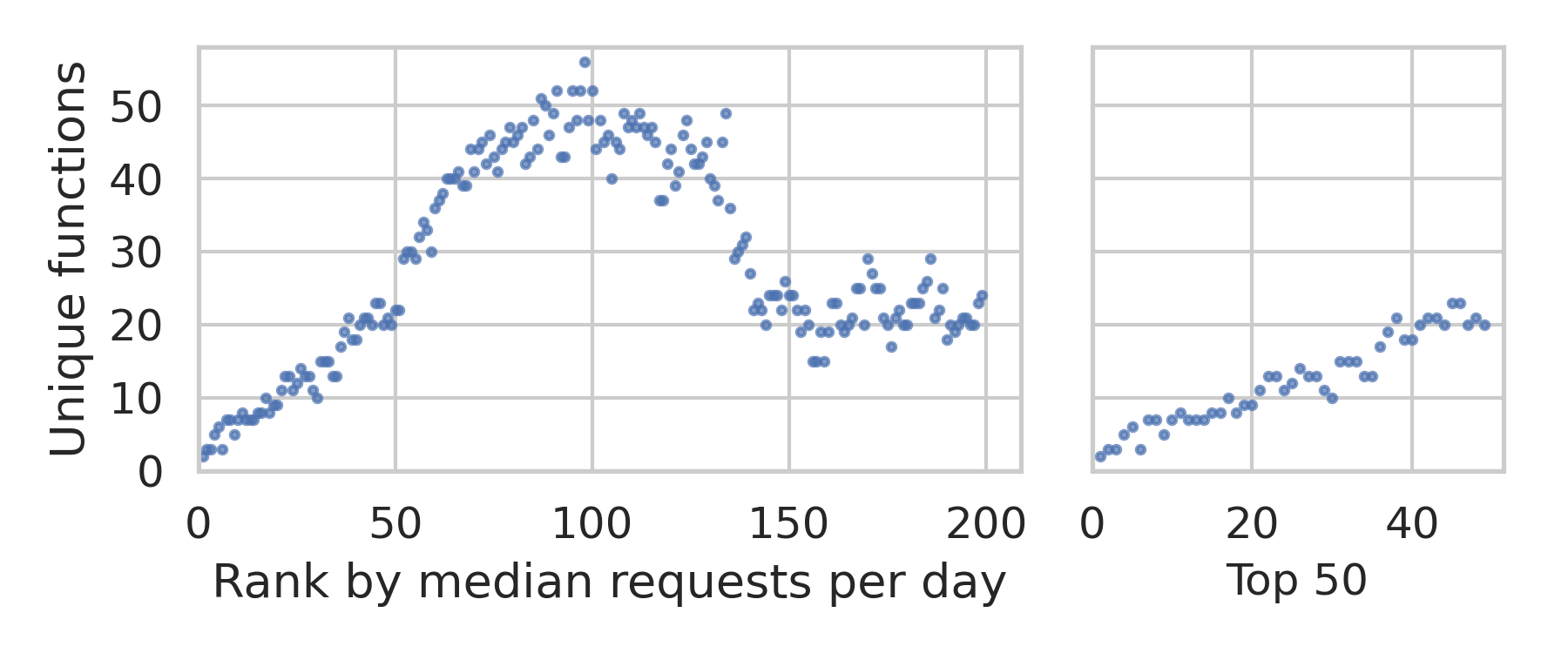}
        \vspace{- 6 mm}
       \caption{Huawei Private.}
       \label{fig:unique_functions_per_rank_wf_full}
    \end{subfigure}
    \begin{subfigure}[b]{0.49\textwidth}
       \includegraphics[width=\linewidth]{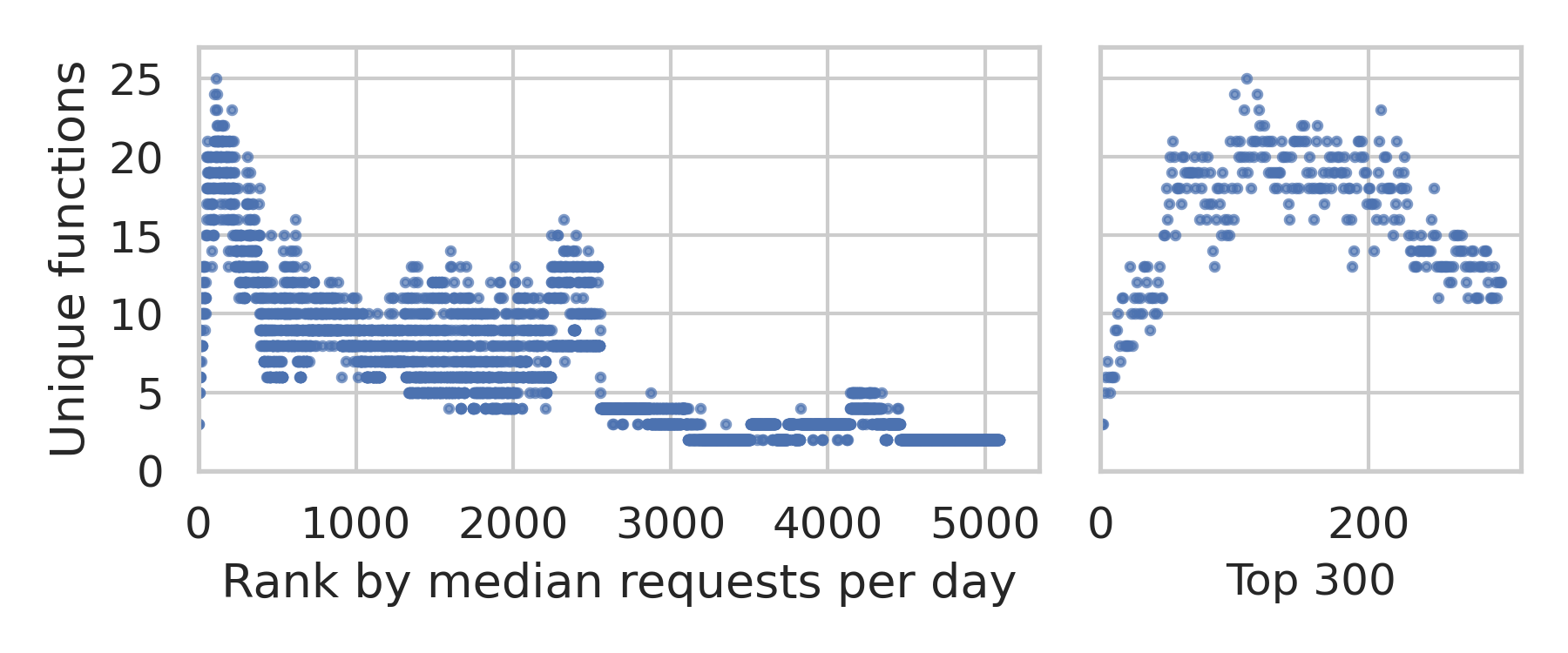}
        \vspace{- 6 mm}
       \caption{Huawei Public.}
       \label{fig:unique_functions_per_rank_fg_full} 
    \end{subfigure}
      \vspace{- 6 mm}
    \caption{Number of unique functions occupying a given rank in Huawei Public and Huawei Private.}
    \label{fig:unique_functions_per_rank}
\end{figure}

To better understand the dynamics over the full duration of the trace, we count the number of unique functions per rank, i.e. number of unique functions per row in the bump chart for the entire duration of the trace, and plot this against the rank (ranked by median), as shown in Figure \ref{fig:unique_functions_per_rank_wf_full} for Huawei Private and Figure \ref{fig:unique_functions_per_rank_fg_full} for Huawei Public. 
Figure \ref{fig:unique_functions_per_rank_wf_full} shows that 2 functions occupy the top spot over the full duration of the Huawei Private dataset. Also, note that the number of unique functions per rank peaks at comparable points in both datasets (around rank 100). 

\begin{tcolorbox}
Higher ranks are occupied by a small number of unique functions. Similarly, low-ranking functions also tend to only have very few variations, with most ranking changes happening in the middle ranks. A scheduler could possibly make use of this fact in deciding which functions can be colocated on the same machines.
\end{tcolorbox}

\subsection{Feature level correlations}

\begin{figure}[!htbp]
	\centering
	\includegraphics[width=1\linewidth]{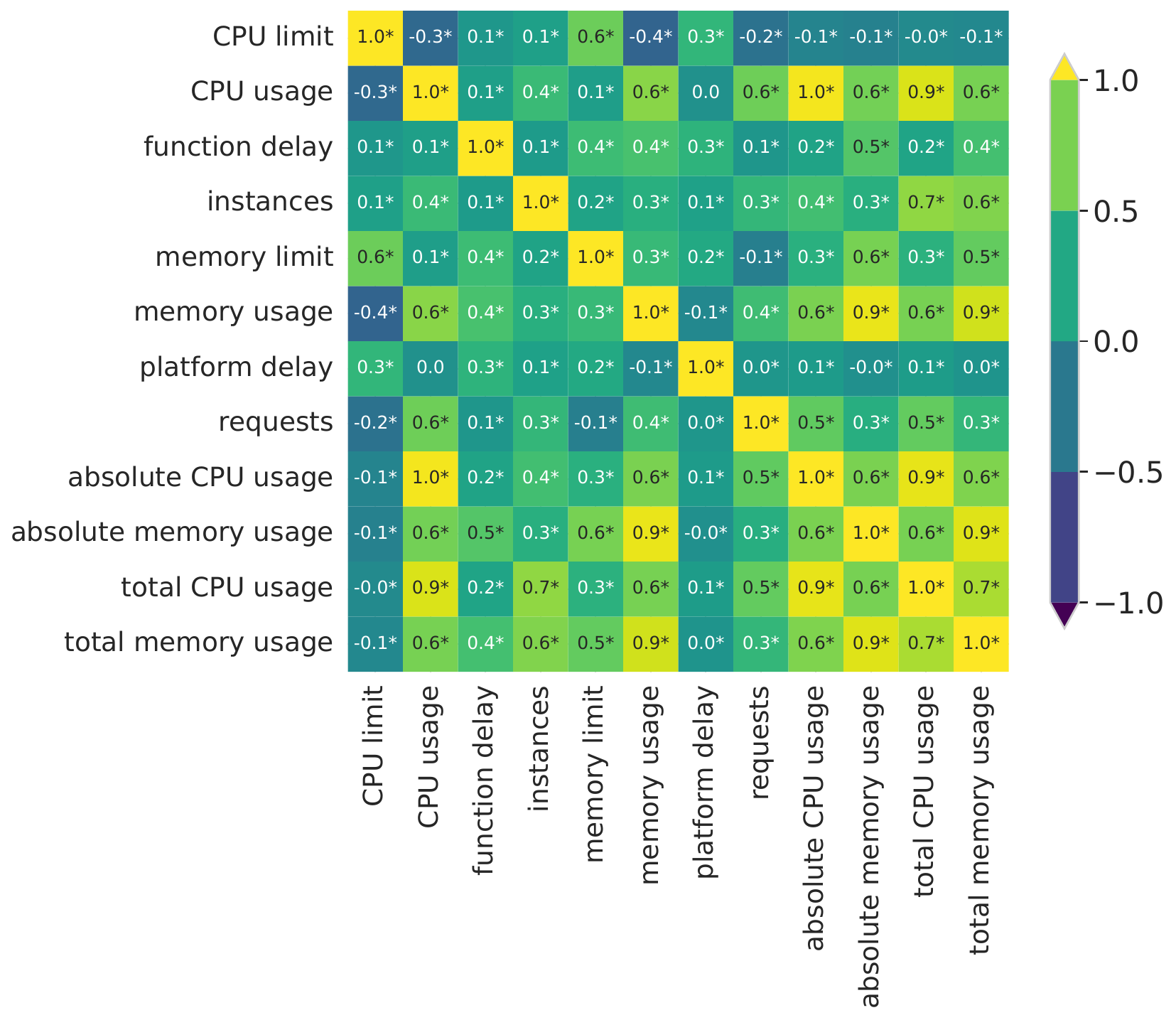}\vspace{-4 mm}
	\caption{Correlation of all metrics averaged over all functions in Huawei Private.}
 \label{fig:correlation_mean_all_functions}
\end{figure}

The Huawei Private dataset has nine metrics (fields) as shown in Table~\ref{tab:datasets}. We have added two additional measures of CPU and memory usage: absolute usage (percentage usage multiplied by the limit, so the result is in MB or cores), and total usage (absolute usage multiplied by the number of pods, representing the total memory or CPU used by all pods running this function).
Figure~\ref{fig:correlation_mean_all_functions} shows a heatmap of the correlations between the features for all the Huawei Private functions. The heatmap shows some interesting insights. First of all, both the function and platform delays are only very weakly correlated with any of the other metrics. In addition, the number of requests is only very weakly correlated with any of the other metrics, if at all. There is, however, a correlation between the memory and CPU metrics.

\subsection{Resource Usage}

\begin{figure*}
	\centering
         \begin{subfigure}[b]{0.38\textwidth}
            \centering
        \includegraphics[height=6.5cm]{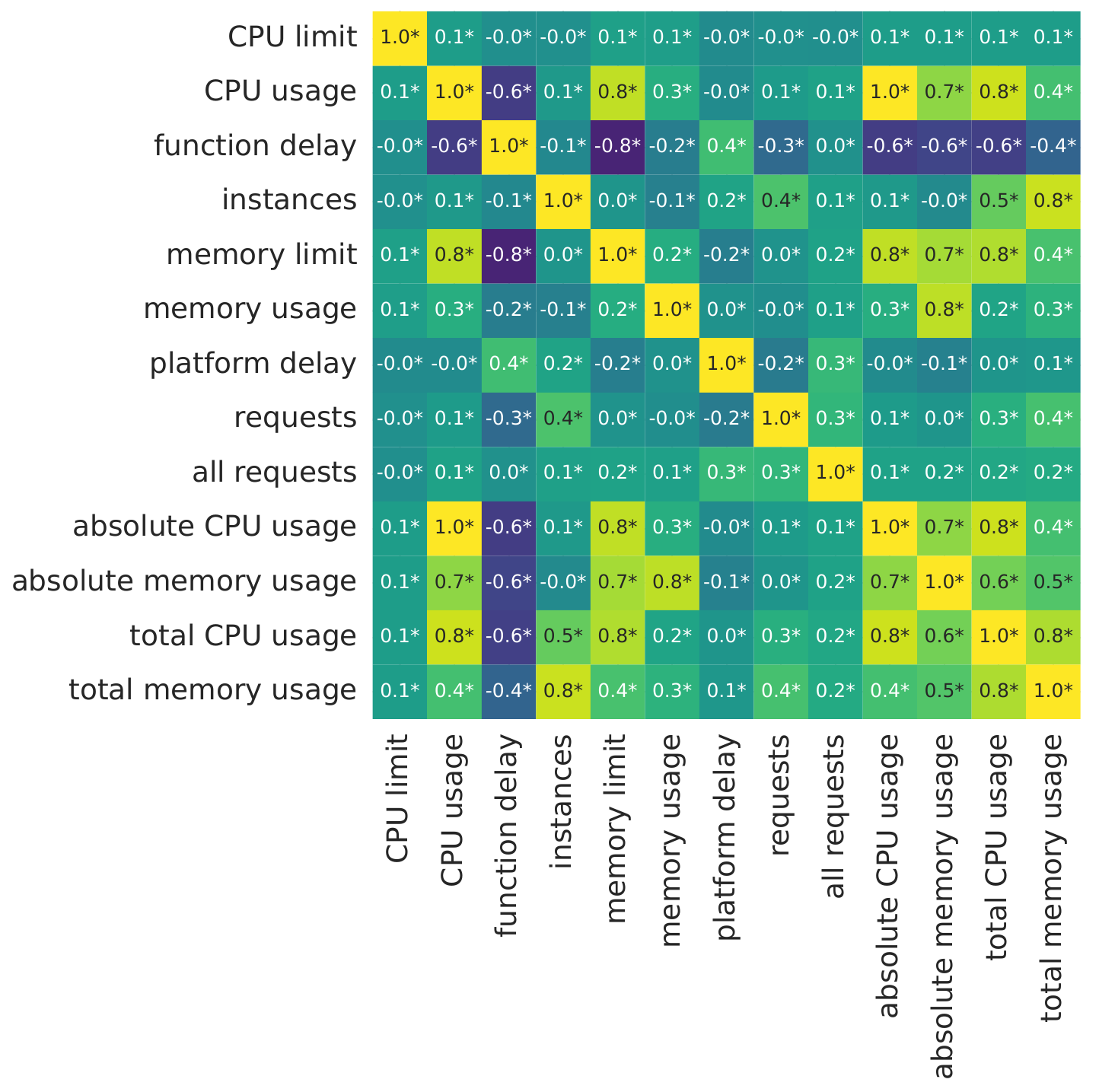}\vspace{-3mm}
        	\caption{Function 66.}
         \label{fig:correlation_66}
         \end{subfigure}
         \hfill
         \begin{subfigure}[b]{0.3\textwidth}
        	\centering
        	\includegraphics[height=6.5cm]{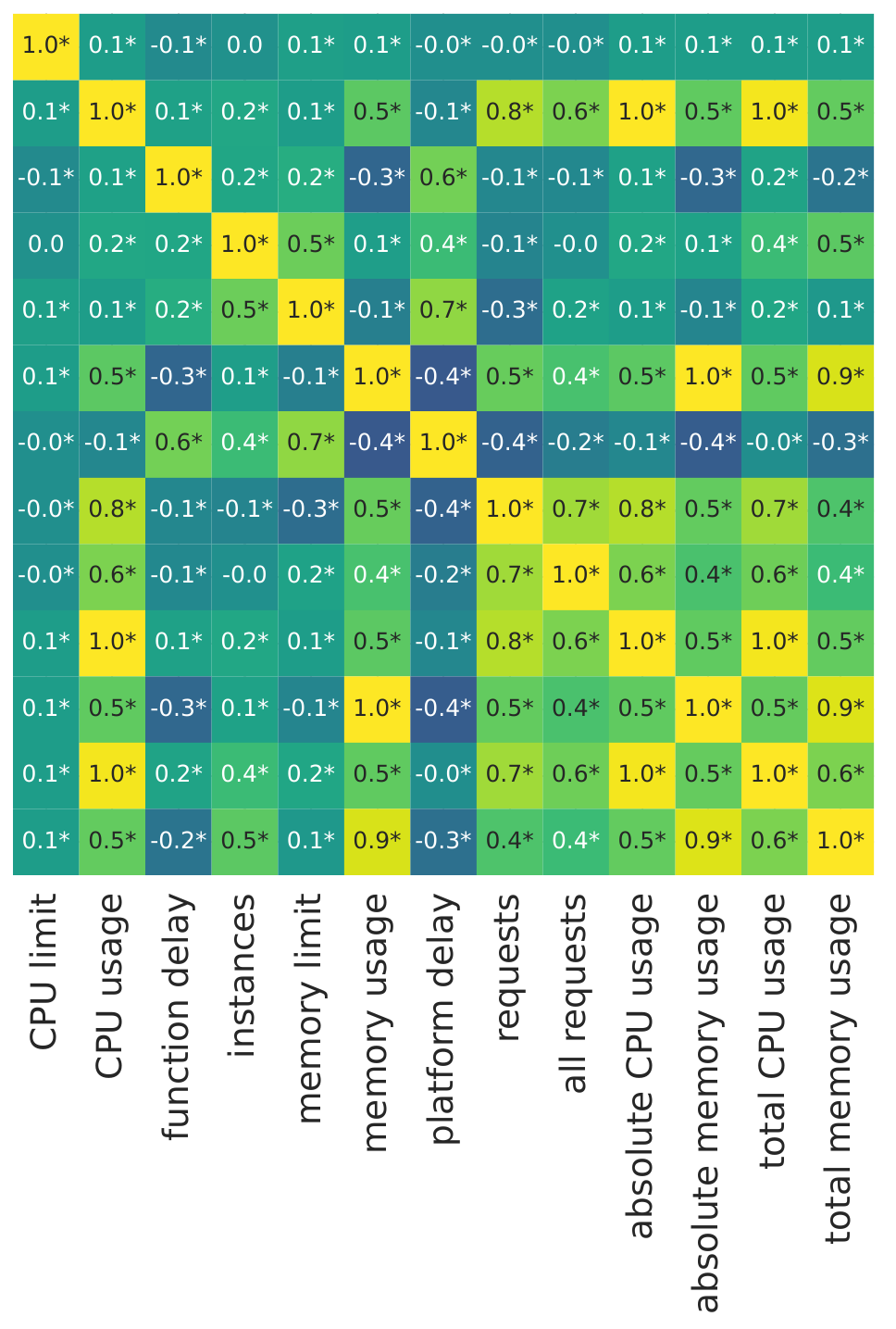}\vspace{-3mm}
        	\caption{Function 72.}
         \label{fig:correlation_72}
         \end{subfigure}
        \hfill
         \begin{subfigure}[b]{0.3\textwidth}
        	\centering
        \includegraphics[height=6.5cm]{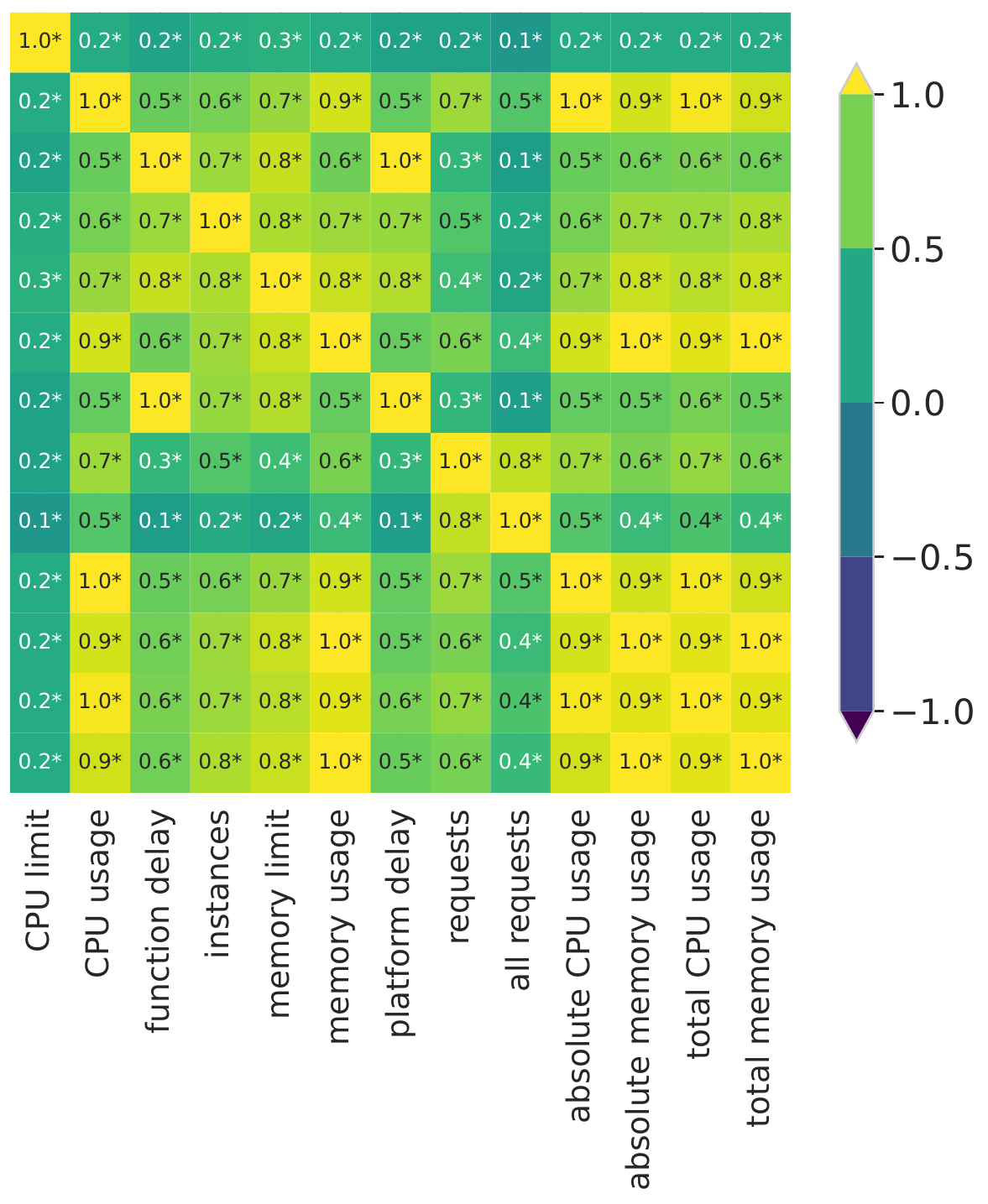}\vspace{-3mm}
        	\caption{Function 75.}
         \label{fig:correlation_75}
        \end{subfigure}
                      \vspace{-2 mm}
 \caption{Correlations for three sample functions in Huawei Private.}
 \label{fig:example_function_correlations}
\end{figure*}

Investigating feature-level correlations for individual functions, Figure \ref{fig:example_function_correlations} shows the Spearman correlation heatmap between all features of three example functions from the top 10 most invoked from Huawei Private. Starting with function 72 in Figure \ref{fig:correlation_72}, we see that for some metrics, there seem to be slightly stronger correlations compared to the averages in Figure~\ref{fig:correlation_mean_all_functions} such as between platform delay and memory limit, or between number of requests and resource consumption. These correlations are very strong in function 75 in Figure~\ref{fig:correlation_75}, with high correlations between function and platform delay and most other metrics. Additionally, for function 66 in Figure \ref{fig:correlation_66} we see negative correlation between function delay and a number of other metrics.

\subsection{Inter-function correlations}

\begin{figure}
     \centering
     \begin{subfigure}[b]{0.22\textwidth}
         \centering
         \includegraphics[height=3.5cm]{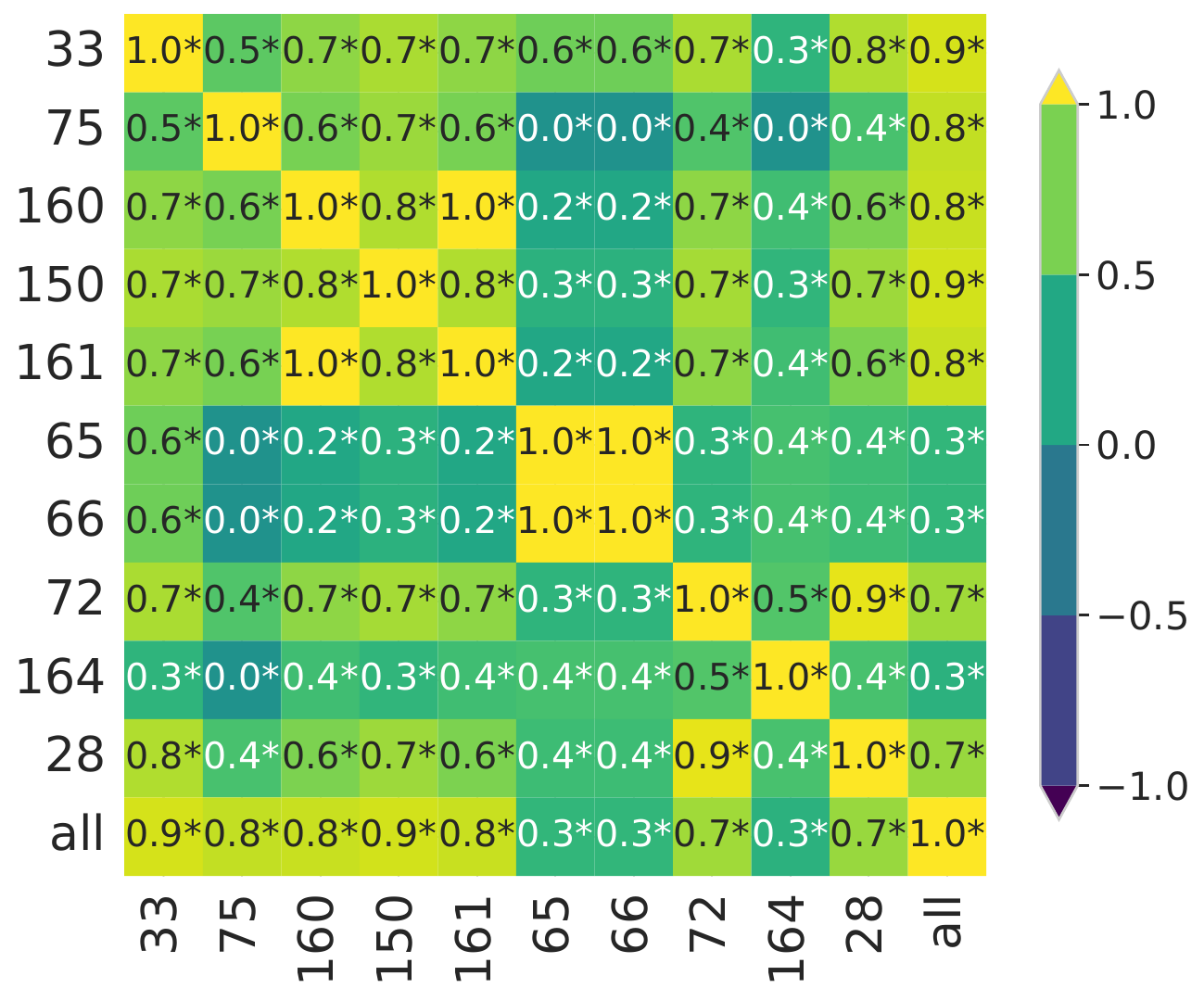}
                \vspace{- 5 mm}
         \caption{Huawei Private.}
         \label{fig:correlation_wf}
     \end{subfigure}
     \hfill
     \begin{subfigure}[b]{0.22\textwidth}
         \centering
         \includegraphics[height=3.5cm]{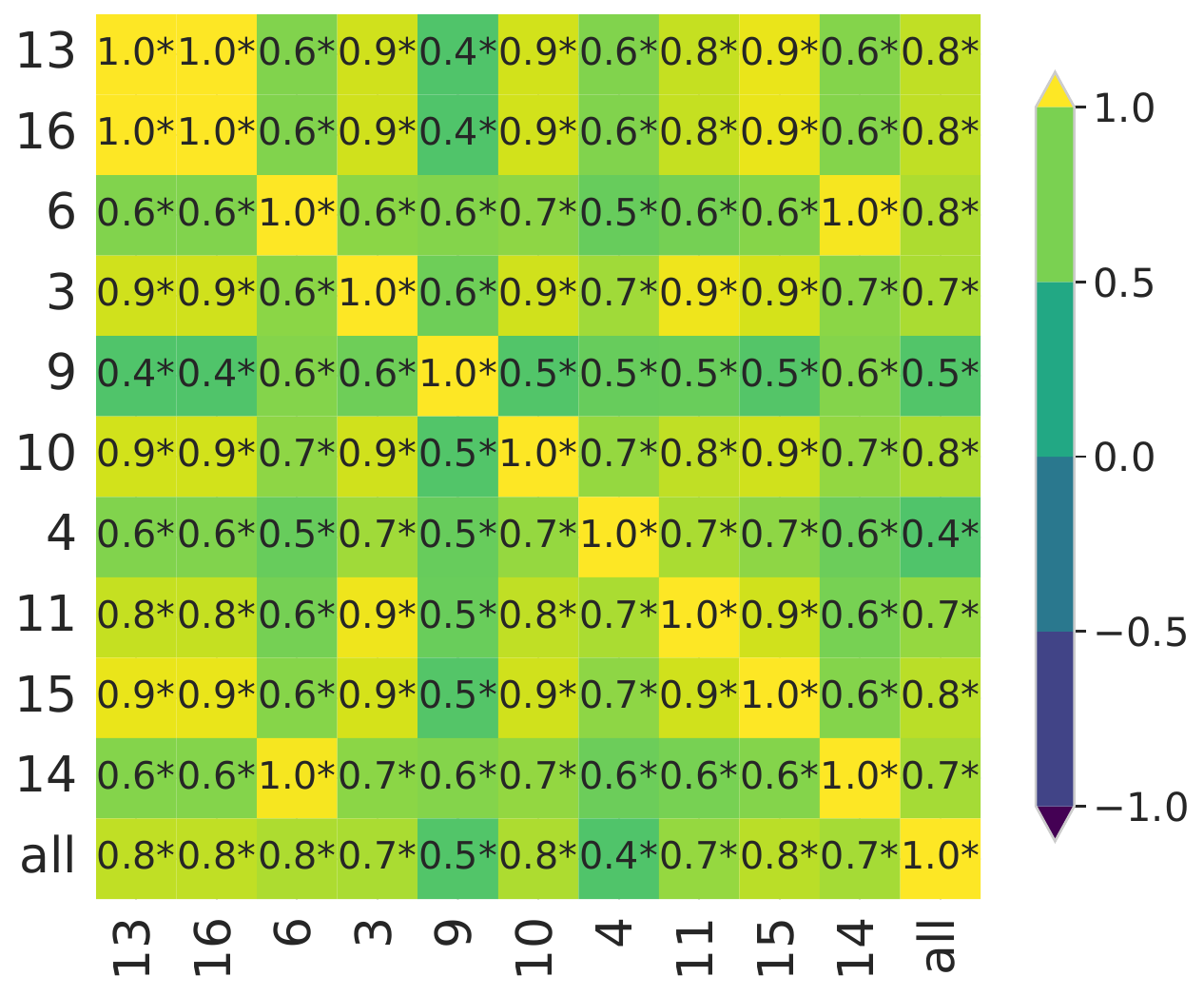}
                \vspace{- 5 mm}
         \caption{Huawei Public.}
         \label{fig:correlation_fg}
     \end{subfigure}
       \vspace{- 4 mm}
        \caption{Correlation of requests of top 10 functions and all requests using Spearman correlation.}
        \label{fig:correlations_full_dataset}
\end{figure}

Since a data center hosts a variety of workloads, it is important for operators to know if there are strong correlations between running workloads. For example, knowing if different function invocations have aligned peaks may inform scheduling. Functions can have correlated bursts because of invocation chaining, or the diurnal pattern of users. 

Figures \ref{fig:correlation_wf} and \ref{fig:correlation_fg} show the correlation heatmap of the top ten functions by invocations for both traces respectively. Each element in the correlation matrix is labeled with its Spearman correlation value, with an asterisk next to values where $p<0.05$. For both datasets, we see very strong correlations between many of the top ten functions, with some functions almost perfectly correlating with each other, suggesting that they burst together. Since the top ten functions by invocations can have billions of invocations per day, this means that from a resource management point of view, it is important to understand which functions burst together to be able to allocate enough (warm) resources to reduce or alleviate altogether the cold start-problem.

\begin{tcolorbox}
    Correlated bursts are common within serverless workloads. Understanding these correlations is important to build scalable serverless resource management systems.
\end{tcolorbox}

\subsection{Workload Granularity and Burstiness}

\begin{figure}
	\centering
	\includegraphics[width=0.95\linewidth]{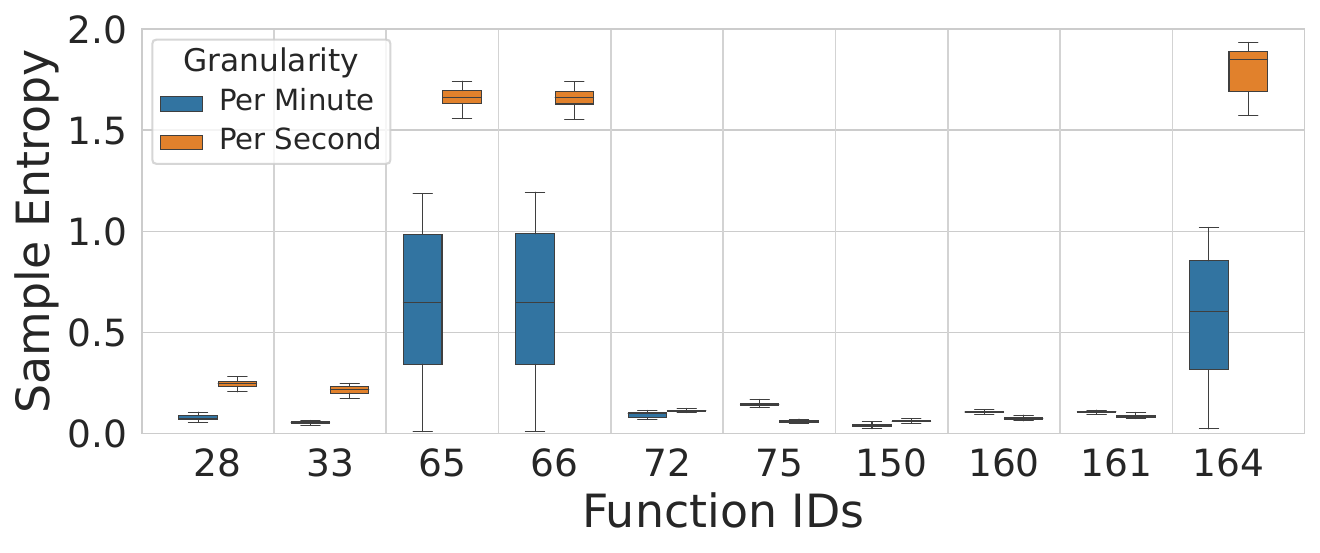}\vspace{-4 mm}
	\caption{Sample entropy box plot for top 10 functions in Huawei Private for different time granularities.}
 \vspace{- 4 mm}
 \label{fig:boxplot_se}
\end{figure}

One important aspect that needs to be taken into account when building serverless, and cloud management systems in general, is the granularity of data logging. On the one hand, very fine-grained logging increases logging costs. On the other hand, coarse-grained logging may introduce significant changes to some statistical aspects of the data. In this Section, we chose burstiness as a measure of statistical variation due to data logging granularity and demonstrate its effects on arrival requests prediction as a use case.

\textbf{Measuring burstiness.} To measure how bursty a workload is, several measures and approaches have been proposed in the literature~\cite{ali2014measuring}. For our analysis, we chose \emph{sample entropy}~\cite{richman2004se} which is a robust burstiness measure, first developed to classify abnormal (bursty) physiological signals, but has since been used to quantify complexity in network traffic~\cite{ali2014measuring,riihijarvi2009measuring,petkov2013characterizing}. Sample entropy is defined as ``the negative natural logarithm of the (empirical) conditional probability that sequences of length $m$ similar point-wise within a tolerance $r$ are also similar at the next point''~\cite{richman2004se}. For our workloads, calculating sample entropy effectively compares sliding windows of length $m$ of the workloads checking if they have any deviations in the number of invocations larger than $r$. We use Python's Antropy library~\cite{Antropy} for our burstiness experiments.

Figure~\ref{fig:boxplot_se} shows a box plot of the average per-day sample entropy for the top 10 functions in our Huawei Private dataset when using per-second versus per-minute granularity. For the per-minute data, we use the sum of invocations for a function over a minute and do not average it.  The plot shows that for Functions $65$, $66$, and $164$, the per-second granularity shows increased burstiness compared to the per-minute granularity trace. For other functions, the difference is insignificant, or much smaller, suggesting that the aggregation did not affect that function's burstiness significantly. For three functions, namely $75$, $160$, and $161$, the burstiness increases slightly due to aggregation. This is a side-effect of using the \emph{sum} over a minute rather than the \emph{mean} over a minute since small anomalies can add up during the minute to magnify into a larger burst. 
\begin{tcolorbox}
    Our analysis confirms that per-second aggregation reveals more trends compared to per-minute aggregation. Generally, it is important to accurately predict per-second dynamics to be able to adapt apriori to bursts. Per-minute aggregation can lead to sub-par or inaccurate resource management decisions.
\end{tcolorbox}

\section{Forecasting challenges}\label{section:forecasting}

Function request arrival forecasting can be used in combination with an estimate of function execution time to estimate in advance the amount of resources allocated for serving a particular function at a given time. This can mitigate the cold start issue, where demand exceeds allocated resources at that time. Ideally, such predictions should be fine-grained to exploit fine-grained patterns, especially in functions with short execution times. 
Fine-grained forecasting is important for multiple internal resource management use-cases. For example, per-second forecasting can be used for resource over-allocation for the majority of functions as they have runtimes of a few milliseconds, as shown in Figure \ref{fig:cdf_platform_delay}. 

In this Section, we study and evaluate several time series forecasting models on the fine-grained long-term (FGLT) data described in this paper. We focus on the top 10 functions. Our objective is to demonstrate the challenges of forecasting FGLT data rather than provide a comprehensive solution.

Forecasting is important to cloud performance optimization (e.g. for minimizing cold starts in FaaS). We focus on forecasting function request arrivals as it is a common use-case in cloud systems. The concept of `fine-grained' data is relative; \textbf{in our case, we consider `fine-grained' to be forecasting tasks where the data sampling rate is much higher than the strongest periodicity}.
For example, Figure \ref{fig:correlogram} shows that the strongest periodicity in our data is every 24 hours, while the sampling rate is 86400 samples per day (the number of seconds in a day). 
Our study illustrates challenges common to all FGLT cloud data as listed below.

\begin{figure*}[!htbp]
    \centering
     \includegraphics[width=0.95\linewidth]{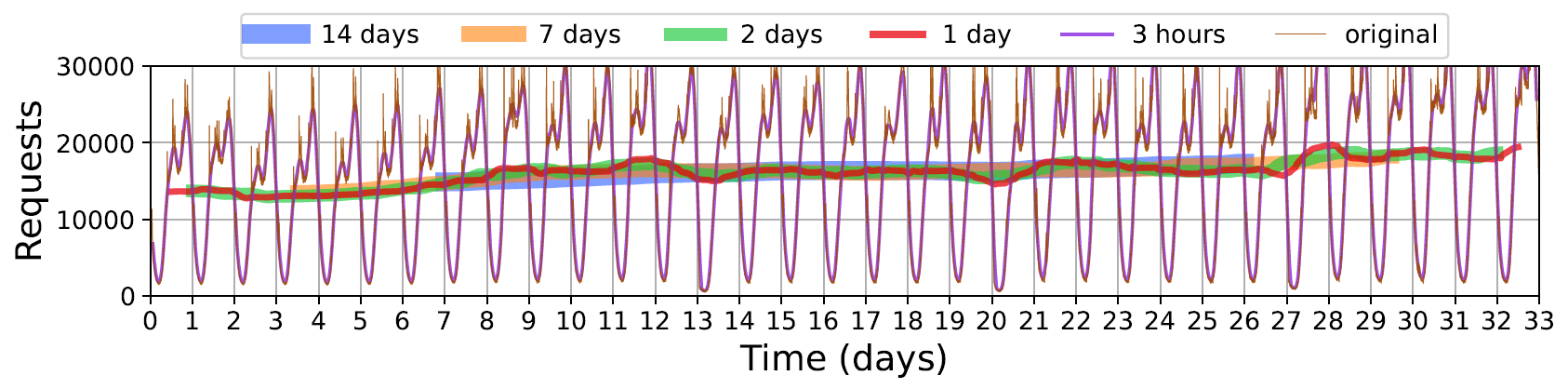}\vspace{-5 mm}
	\caption{A 33 day chunk of an example function in Huawei Private. The different color lines show different scale trends computed as windowed means of the data with various window sizes (from 14 days to 3 hours).}
 \label{fig:sliding_window_fn_id_93}
\end{figure*}

\begin{enumerate}
    \item \textbf{Fine-grained data is expensive for powerful models to ingest and utilize.} For a model to ingest one week of per-second data, it needs to ingest and learn long and short term patterns from $86400 \times 7=604,800$ data points. However, ingesting this amount of data is infeasible in most practical cases for most models because of the difficulty of learning such large feature spaces. That is, models have to learn fine-grained minute or second level patterns while also accounting for long-term trends over days or weeks, as shown in Figure \ref{fig:sliding_window_fn_id_93}. As of yet, no models are designed for this paradigm. See Section \ref{sec:exp-second-challenges} for more information.
    \item \textbf{Standard univariate or multivariate models are poorly suited to function request forecasting} since for each new function we want to predict, the model needs to be either retrained or tuned, which is too expensive.
    This is discussed further in Section \ref{sec:setup}.
    \item \textbf{Long-term forecasting} Our system runs tens to hundreds of thousands of functions. Running a prediction for each of these functions frequently is very expensive and the overhead of forecasting can quickly outweigh the utility of predictions. Therefore, we expect a forecast to be relevant for at least one day, such that the model is only queried once per day.
\end{enumerate}

\subsection{Experimental setup}\label{sec:setup}
\noindent\textbf{Models.}\label{sec:models}
We trained eight models on the top 10 functions for both platforms: linear regression, TimesNet \citep{wu2022timesnet}, N-HiTS \citep{challu2022n}, FFT extrapolation, NeuralProphet \citep{triebe2021neuralprophet}, PatchTST \citep{PatchTST2023}, TiDE\cite{TiDEdas2023longterm}, and N-Linear \citep{zeng2022transformers}. We describe these models later in detail. Where possible, we trained global univariate models~\cite{herzen2022darts} when the original model supports this training regime.
\emph{Global univariate models} are models trained on samples from multiple time series. This training regime produces improvements for some scenarios~\cite{herzen2022darts, global_models_hansika_2020, Konstantinos_survey_2022} over models that are only trained on samples from the same time series they are trying to predict (local models). Another advantage is that training a global model allows us to predict other single univariate time series that it was not necessarily trained on. 

Global univariate models are well-suited to data in cloud infrastructure because:
\begin{enumerate}
    \item Global univariate models are typically used where time series are similar or related. Functions often exhibit similar patterns, meaning that a function-specific model is not necessary.
    \item Thousands of functions may require forecasting, so function-specific models are too costly for both training and inference.
    \item Multivariate models (where inputs are variables from multiple functions) are not feasible for thousands of functions. One may consider narrowing down the input to the top-$k$ functions, but the top-$k$ ranking changes over time (see Figure \ref{fig:bump_charts}) and entirely unseen functions may appear.
\end{enumerate}

In our work, only two models do not support global univariate training; NeuralProphet and FFT. These local models are instead fitted to each function individually. We include these two models as they are simple and lightweight models that are widely used. However, they suffer from all the downsides of local models mentioned before. 

We now describe the eight models we train in more detail:
\begin{enumerate}
    \item \textbf{FFT Extrapolation} computes the FFT of the input and extrapolates future values. In our implementation, data is de-trended with a third-degree polynomial and the top 100 frequencies are kept for extrapolation.
    \item \textbf{Linear regression} is a simple forecasting model that assumes a linear relationship between output forecasts over the forecast horizon and input features taken from previous values in the sequence.
    \item \textbf{N-Linear} is a simple linear model with an additional preprocessing step to account for distributional shift in the data.
    \item \textbf{NeuralProphet} combines classical forecasting and deep learning, consisting of trend, seasonality, and residual components. The autoregressive component is disabled because it would require many invocations per day, which is not permitted in our use case.
    \item \textbf{N-HiTS} uses multi-rate signal sampling and hierarchical interpolation to learn patterns at different scales, and then recombine these components into an overall forecast. We use 3 stacks, 1 block per stack and 2 fully connected layers preceding the final forking layers in each block of every stack. Each layer has width 512.
    \item \textbf{TimesNet} uses a multi-scale approach that converts 1D series to 2D and then uses convolutional kernels. In experiments, we use 2 encoding layers, 1 decoding layer, model dimension of 64.
    \item \textbf{PatchTST} is an encoder-only transformer architecture that uses patching, where a series in split into subsequences before passing through the model, and channel independence, where different channels are predicted independently from each other, but sharing model weights while training. We use a patch length of 16, 2 encoding layers and model dimension of 64.
    \item \textbf{TiDE} is a multi-layer perceptron (MLP) based encoder-decoder model specifically designed for long-term time series forecasting.
\end{enumerate}

\noindent\textbf{Data Description.} For training, we selected the top 10 functions by median requests. For Huawei Private, we train models on 38 days consisting of two chunks with a gap in between. We use the following 7 days for validation to implement early stopping (to mitigate overfitting) and test on the next 7 days. For Huawei Public, we train on the first 20 days, use the following 3 days for validation, and test on the next 3 days.

We study the performance of the eight models on per-minute and per-second data. We aggregate Huawei Private data per minute. For training a global model for per-minute forecasting, we use an input length of two days for Huawei Private and one day for Huawei Public. We set the forecast horizon to one day, i.e., we predict an entire day in advance. This is closest to how the model would be used in production. For training a global model for per-second forecasting (N-HiTS only) we use a 6-hour input length and a one-hour forecast horizon as most models fail with longer inputs.

\looseness=-1

\subsection{Forecasting results}\label{sec:exp-forecasting}

Table \ref{tab:datasets} shows the results of our forecasting models on Huawei Public, and Huawei Private at per-minute and per-second resolutions. We show root-mean-squared error (RMSE), mean absolute error (MAE), and mean absolute percentage error (MAPE). We give results on the top 5 functions due to space constraints. In the table, we highlight the best performing model for each function's workload in boldface. In many cases, an FFT extrapolation performs better than more complex methods. Per-second data disables the use of many models since they cannot ingest enough data to meet our long-term requirement (i.e., forecast for a full day).

\begin{figure}[!htbp]
    \centering
    \begin{subfigure}[b]{0.49\textwidth}
       \includegraphics[width=\linewidth]{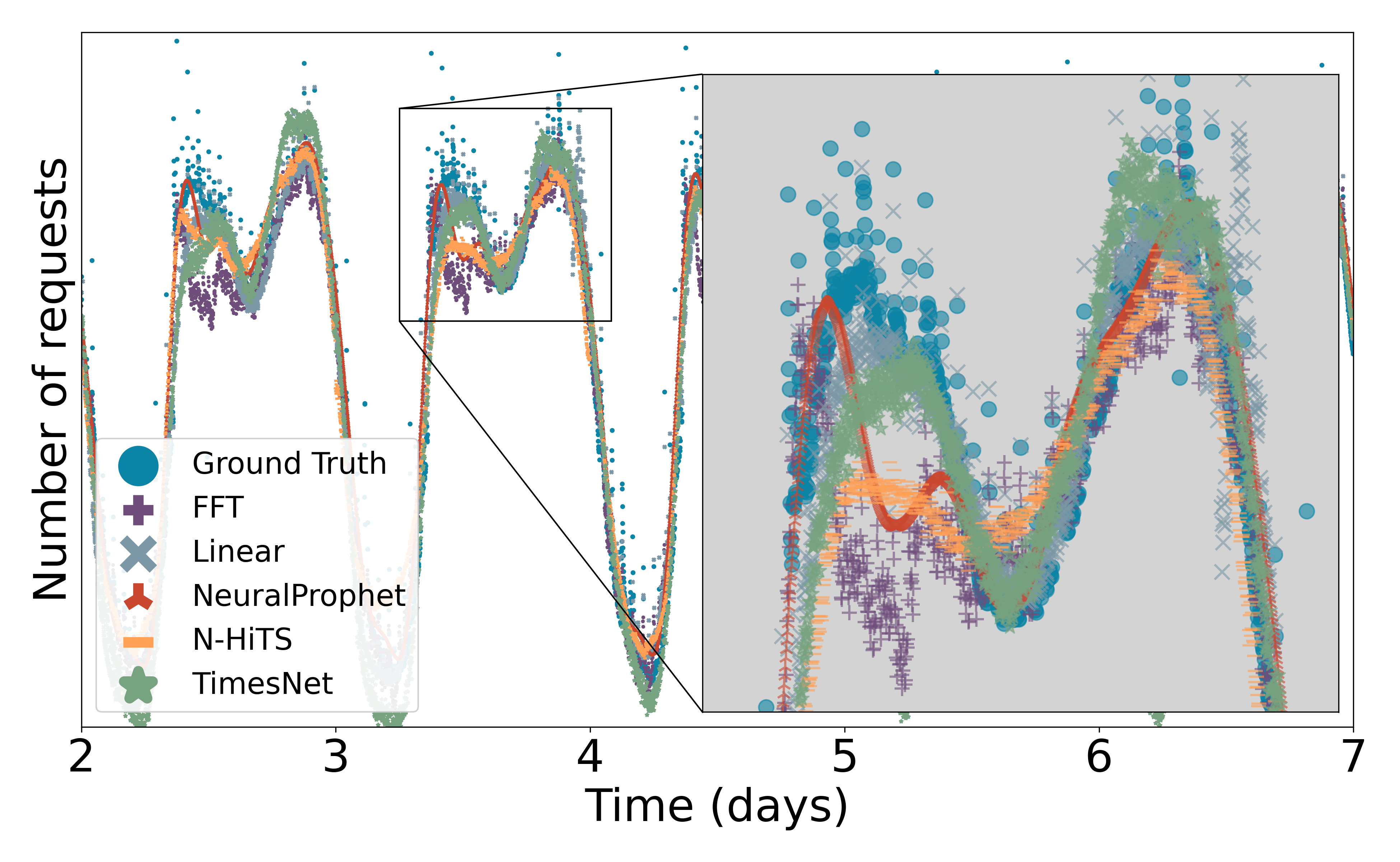}
       \vspace{- 8 mm}
       \caption{Per-minute experiments.}
       \label{fig:example_predictions_minutes} 
    \end{subfigure}
    
    \begin{subfigure}[b]{0.49\textwidth}
       \includegraphics[width=1\linewidth]{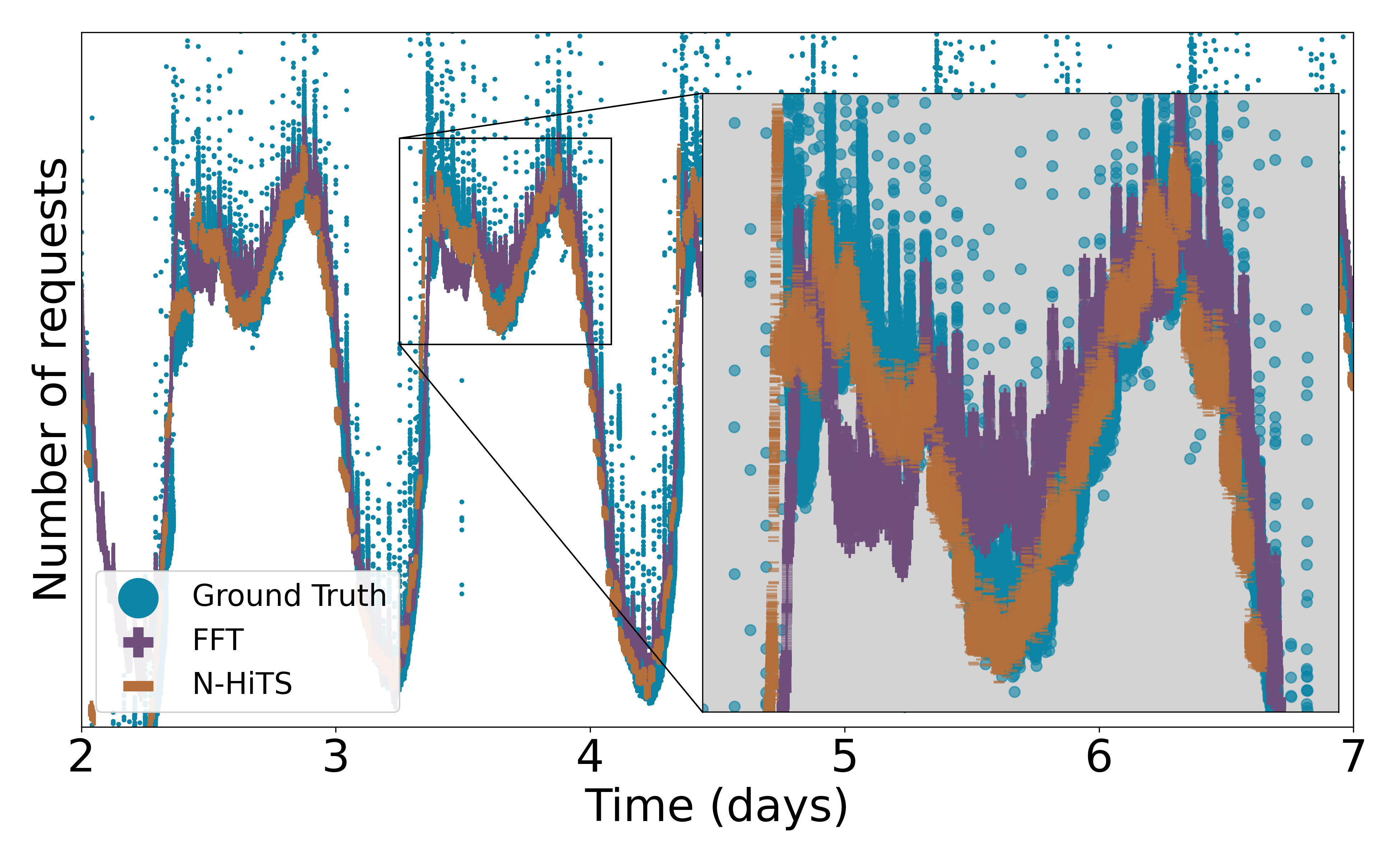}
              \vspace{- 8 mm}
       \caption{Per-second experiments.}
       \label{fig:example_predictions_seconds}
    \end{subfigure}
           \vspace{- 6 mm}
    \caption{Example function request predictions for different machine learning models.}
    \label{fig:example_predictions}
\end{figure}

Figure \ref{fig:example_predictions} visualizes the results from Table \ref{tab:datasets}. Figure \ref{fig:example_predictions_minutes} shows predictions for per-minute data, where most models make reasonable predictions of large-scale trends. However, the zoomed-in view shows that models fail to accurately predict finer-grained patterns. Figure \ref{fig:example_predictions_seconds} shows predictions for per-second data. Again, models struggle with finer-grained patterns. For both per-minute and per-second forecasting, the zoomed-in plots in Figures \ref{fig:example_predictions_minutes} and \ref{fig:example_predictions_seconds} show that predictions deviate significantly from ground truth data for several hours.
For per-second forecasting, many models run out of memory or require prohibitively long training, so only FFT and N-HiTS could be evaluated.

\begin{figure}
	\centering
	\includegraphics[width=0.85\linewidth]{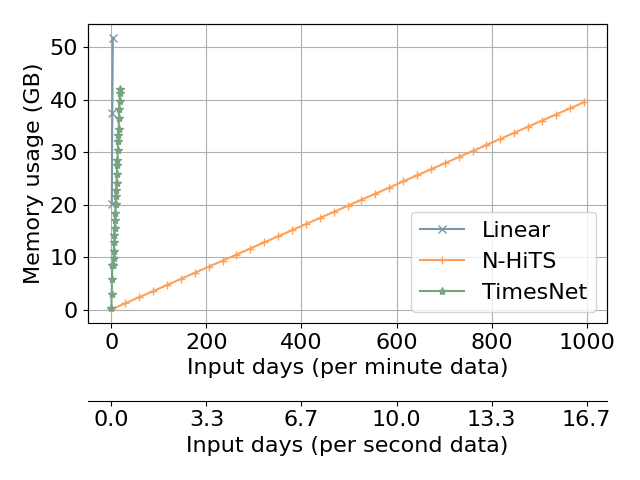}\vspace{-6 mm}
	\caption{Memory consumption of selected models.}
 \label{fig:memory_footprint}
\end{figure}

To better understand the scalability problems of complex prediction models, Figure \ref{fig:memory_footprint} shows allocated memory of selected models with the number of input days during training. Note that N-HiTS scales best, while linear scales worst. 

While it was not possible to train N-HiTS with the standard one-day-in-one-day-out setup due to memory and training time constraints, we train N-HiTS with a six-hour input length and one-hour output length as shown in Figure \ref{fig:example_predictions_seconds}. However, this increases the cost of predicting all functions in our data centers as we now invoke the model 24 times more often. The other model that works with our data is FFT. FFT fits large-scale trends, but does not predict finer-grained patterns. Importantly, Figure \ref{fig:example_predictions} shows all models are poorly suited to predicting peaks, especially for per-second data. More work is required to accurately predict peaks of bursty time series, given the importance of bursts in cloud platforms.

It is worth noting that many simple, non-neural network models perform comparably or sometimes better than state of the art neural networks. It appears that for FGLT forecasting, neural networks give marginal improvements over simpler methods. This is especially noteworthy given the overhead of training and inference of neural networks.
We found N-HiTS to be one of the more robust and performant models. This may be because its architecture is specifically designed for learning patterns at different scales.

\begin{table*}[!htbp]
    \centering
     \includegraphics[width=1.0\linewidth]{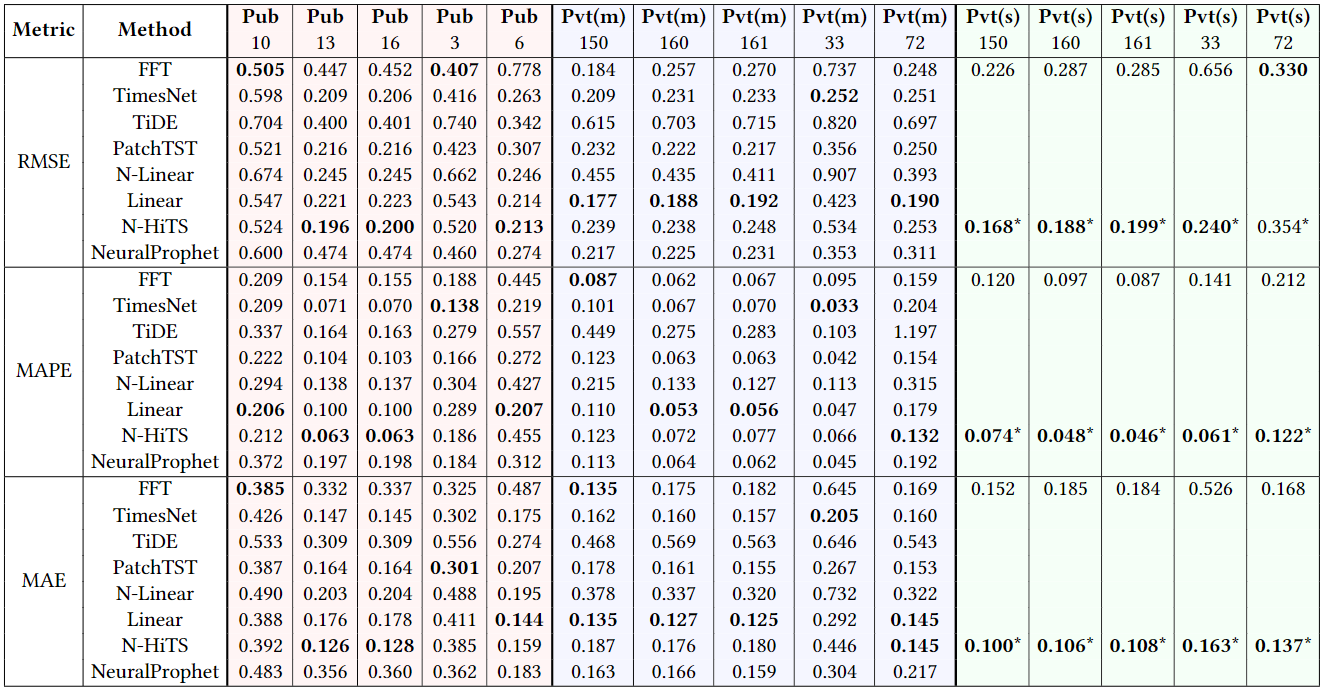}
	\caption{Results for forecasting experiments. For Huawei Public (Pub) results are computed in a rolling manner over 2 days with input of 1 day. For Huawei Private per-minute -- Pvt(m) -- results are computed in a rolling manner over 5 days with an input of 2 days. For Huawei Private per-second -- Pvt(s) -- results are computed using a rolling forecast over 1 hour. The best results are highlighted in \textbf{bold}.  All metrics are computed on normalized data. *For per-second forecasting, N-HiTS uses a six-hour input window one-hour output window as explained in the text.}
 \vspace{-3mm}
 \label{tab:results}
\end{table*}

\subsection{Per-second challenges}\label{sec:exp-second-challenges}
Fine-grained long-term forecasting is under-explored in literature. Even if a model could ingest such amounts of data with trends at multiple scales as shown in Figure \ref{fig:sliding_window_fn_id_93}, it would not necessarily learn the underlying trends. The memory overhead of consuming this much data means that even efficient models such as N-HiTS can only ingest approximately 18 days (see Figure \ref{fig:memory_footprint}).

The per-second results in Table \ref{tab:results} and Figure \ref{fig:example_predictions_seconds} further show the challenge with more fine-grained data. Therefore, we see the following challenges in fine-grained long-term time series forecasting model capabilities; (1) Ingesting long-term per-second data, where long-term means at least 1 week; (2) Generalizing well under the above requirement, without severe overfitting; and (3) Forecasting (directly or autoregressively) long-term without severe degradation.

\section{Related Work}

Workload characterization of server systems has been a popular research topic for decades~\cite{feitelson2015workload,arlitt1996web,crovella1997self}. Recently, there have been a few studies on characterization and analysis of cloud system production workloads, e.g., from Google's Borg clusters~\cite{borg_the_next_generation_2020}, Azure's VM workloads~\cite{ResourceCentral2017}, Alibaba ML workloads~\cite{weng2022mlaas}, and Azure Serverless workloads
\cite{serveless_in_the_wild_2020,mahgoub2022wisefuse,zhang2021faster}, among many others. The main goal of these studies is to design better systems, where the system management is data-driven. 
With the exception of the workloads analyzed from Azure on serverless systems~\cite{serveless_in_the_wild_2020,zhang2021faster,mahgoub2022wisefuse}, 
 existing work on data center workload analysis focuses on CPU and memory utilization, VM types, and other machine and VM level metrics.
To the best of our knowledge, the only large-scale publicly available FaaS dataset are the two available from Microsoft Azure \cite{serveless_in_the_wild_2020,mahgoub2022wisefuse}, which includes two weeks of function request data for tens of thousands of functions. Another dataset is also to be released from Azure that shows the graph structures of FaaS functions~\cite{zhang2021faster}.


\section{Conclusion}


This paper describes two Huawei serverless traces from public and private infrastructure containing 1.4 trillion function invocations, which we open-source to the research community. It provides a comprehensive analysis covering statistical features of our workloads, followed by a longitudinal analysis of periodicity and ranking of functions across our traces. 


Our analysis uncovered several interesting insights which can inform the engineering of future resource schedulers: requests vary by up to 9 orders of magnitude across functions, with some functions being executed over one billion times per day; scheduling time, execution time and cold-start distributions vary across 2 to 4 orders of magnitude and have very long tails; many functions demonstrate strong periodicity; and the highly ranked functions are occupied by only a small number of unique functions. Our analysis also highlights the need for further research and development in estimating resource reservations and time series prediction to account for the huge diversity in how serverless functions behave.

\section*{Acknowledgments}
We would like to thank the YuanRong lab for their valuable collaboration and collecting the data. We would also like to thank Wei Wei for his contributions and communication with YuanRong. We would like to thank Blesson Varghese for his helpful proofreading and edits. Lastly, we would like to thank the reviewers at ACM SoCC for their insightful comments and Yue Cheng for shepherding.

\newpage
\balance
\bibliographystyle{ACM-Reference-Format}
\bibliography{main}

\end{document}